\documentclass[12pt]{article}
\usepackage{epsf}
\newcommand{\eqb}{\begin{equation}}
\newcommand{\eqe}{\end{equation}}
\newcommand{\dmb}{\begin{displaymath}}
\newcommand{\dme}{\end{displaymath}}
\newcommand{\pd}{\partial}

\newcommand{\eab}{\begin{eqnarray}}
\newcommand{\eae}{\end{eqnarray}}
\newcommand{\ra}{\right\rangle}
\newcommand{\la}{\left\langle}
\newcommand{\e}{\mbox{e}}
\newcommand{\be}{\begin{equation}}
\newcommand{\ee}{\end{equation}}

\newcommand{\La}{\Lambda}

\def\lsim{\mathrel{\raise.3ex\hbox{$<$\kern-.75em\lower1ex\hbox{$\sim$}}}}
\def\gsim{\mathrel{\raise.3ex\hbox{$>$\kern-.75em\lower1ex\hbox{$\sim$}}}}
\def\Li2{{\rm Li}_2}

\newcommand{\bmE}{\mathbf E}

\newcommand{\bmx}{\mathbf x}
\newcommand{\bmy}{\mathbf y}

\begin{document}

\begin{titlepage}
\begin{flushright}
MPI-PhT 2002-25 \\
hep-ph/0206201\\
June 2002
\end{flushright}
\vspace{0.6cm}

\begin{center}
\Large{{\bf Reshuffling the OPE: \\Delocalized Operator Expansion}}

\vspace{1cm}

A.~H.~Hoang and R.~Hofmann

\end{center}
\vspace{0.3cm}

\begin{center}
{\em Max-Planck-Institut f\"ur Physik\\ 
Werner-Heisenberg-Institut\\ 
F\"ohringer Ring 6, 80805 M\"unchen\\ 
Germany}
\end{center}
\vspace{0.5cm}

\begin{abstract}

A prescription for the short-distance expansion of Euclidean current
correlators based on a delocalized modification of the multipole
expansion of perturbative short-distance coefficient functions 
is proposed that appreciates the presence of nonlocal
physics in the nonperturbative QCD vacuum. 
This expansion converges
better than the local Wilson OPE, which is recovered in the limit of
infinite resolution. 
As a consequence, the usual local condensates in the Wilson OPE
become condensates that depend on a resolution parameter 
and that can be expressed as an infinite series of local
condensates with increasing dimension. In a calculation 
of the nonperturbative correction to the ground state energy 
level of heavy quarkonia the improved convergence properties 
of the delocalized expansion are demonstrated. 
Phenomenological evidence is gathered
that the gluon condensate, often being the leading nonperturbative
parameter of the Wilson OPE, is indeed a function of
resolution. The delocalized expansion is applied to
derive a leading order scaling relation for 
$f_D/f_B$ in the heavy mass expansion. 

\end{abstract} 

\end{titlepage}

\section{Introduction}
\label{sectionintroduction}

Establishing a connection between universal parameters of the 
nonperturbative QCD vacuum in Euclidean space-time on the one hand and 
an integral over weighted hadronic cross sections on the other 
hand has proven to be a very fruitful idea.\,\cite{SVZ}
The three underlying assumptions of this approach are 
the analyticity of correlators of gauge invariant currents, the existence 
of a {\sl meaningful} (asymptotic) large-momentum expansion of these
correlators in the Euclidean region and the possibility of a dual
description for hadron dynamics in terms of the dynamics of quarks and
gluons. One may wonder whether these assumptions really are all
independent. Given our poor analytic understanding of the
nonperturbative quantum dynamics of quarks and gluons at present, we
are not in a position to prove the analyticity assumption in QCD from
first principles. On the other hand, one may, on general grounds and
with the help of results from lattice gauge theory, try to connect the
issue of a meaningful large-momentum expansion with that of
quark-hadron duality. 

The conventional approach to the large-momentum expansion of the
correlator of a gauge invariant QCD current $j(x)$ is Wilson's operator 
product expansion (OPE)\footnote{For the purpose of this introduction
it is not necessary to consider the Lorentz and flavor structure of the
current. Also, we assume that $j(x)$ has no anomalous dimension. In general, 
the separation between high and low momenta 
in the OPE is ambiguous, and one has to introduce a factorization 
scale $\mu$ which $c_{Nl}$ and $O_{Nl}$ depend upon. This 
dependence cancels in the whole series.},
\begin{equation}
\label{Wilson}
i\int d^4x\,\e^{iqx}\la\, T j(x)j^\dagger(0)\,\ra\, =\, 
Q^2\sum_{N=0}\sum_{l=1}^{l_N}\,c_{Nl}(Q^2)\,\la O_{Nl}(0)\ra
\ ,\ \ (Q^2=-q^2>0)\ .
\end{equation}
In Eq.\,(\ref{Wilson}) the contribution with $N=0$ is solely due to 
perturbation theory, whereas $N>0$ runs over the dimension of the 
local and gauge invariant 
composite operators $O_{Nl}(0)$, and $l$ labels operators of the same
dimension. Operators $O_{N_1l_1}(0)$ and $O_{N_2l_2}(0)$ with
$N_1<N_2$ need not be unrelated. Rather, $O_{N_2l_2}(0)$ may be the
result of gauge-covariant differentiations of the  gauge invariant
$k$-point function corresponding to the content of 
fundamental fields in the local
operator $O_{N_1l_1}(0)$.\,\cite{SVZ} In this sense $O_{N_2l_2}(0)$ 
is a reducible operator. It is the purpose of this
paper to exploit the relation of a chain of reducible local operators
with nonlocal vacuum averages for constructing a modified version of
the OPE. 

According to the standard interpretation~\cite{SVZ}, the physics
of low-momentum, nonperturbative vacuum fluctuations resides in the
vacuum averages $\la O_{Nl}(0)\ra$, whereas high-momentum,
perturbative propagation is contained in the Wilson coefficients
$c_{Nl}(Q^2)$. If $\la O_{Nl}(0)\ra\sim \La^N$ then,  
apart from logarithmic factors arising from anomalous, perturbative
operator scaling, we have $c_{Nl}(Q^2)\sim Q^{-N}$. Here it is assumed
that $Q$ is the only relevant short-distance scale. Nonperturbative 
corrections to the $N=0$ contribution are 
counted in powers of $\La/Q$ and are small if $\La\ll Q$, and
if $l_N$ is sufficiently small. It is expected
that the OPE is {\sl asymptotic} at best, i.e.\  convergence is only
apparent up to some critical value of $N$.  

The expansion into vacuum averages of {\sl local} operators assumes  
correlations in the vacuum to be rigid, or in other words, it is 
assumed that the lower bound on the correlation lengths of the 
corresponding {\sl nonlocal} gauge
invariant $k$-point functions is considerably larger than the
short-distance scale $Q^{-1}$. More specifically, one likes to think of 
correlation lengths being of the order $\La^{-1}\sim\Lambda_{\rm QCD}^{-1}$, 
where $\Lambda_{\rm QCD}$ is the typical hadronization scale, i.e.\,it
is commonly assumed that there is no hierarchy in 
nonperturbative scales. 

\begin{figure}
\begin{center}
\leavevmode
\epsfxsize=9.cm
\leavevmode
\epsffile[80 25 614 394]{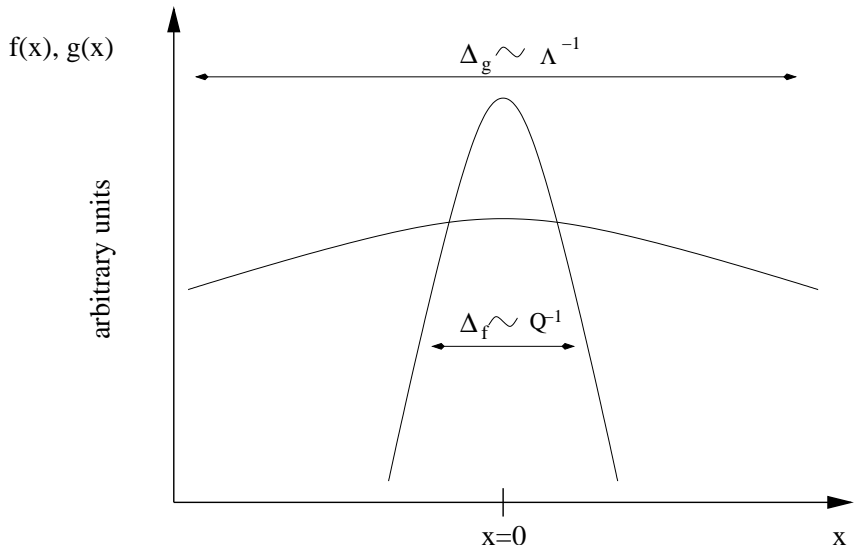}
%
\end{center}
\caption{Schematic drawing of the short-distance function $f(x)$ and
the 2-point-correlator $g(x)$ illustrating the scale hierarchy
$\Delta_f/\Delta_g\sim\Lambda/Q\ll 1$.
\label{figfandg}
}   
\end{figure}

Given the exact knowledge of all gauge invariant QCD vacuum correlators and
maintaining the possibility to factorize perturbative propagation from 
nonper\-tur\-ba\-tive background fluctuations for each operator structure,
one may view the OPE~(\ref{Wilson}) as a 
degenerate multipole expansion of the perturbative part. To see this,
we consider the simplified case of a nonperturbative and nonlocal
dimension-4 structure corresponding to a {\sl slowly varying} 2-point
correlator $g(x)$ falling off at distance $\Delta_g\sim \Lambda^{-1}$ in its 
Euclidean space-time argument $x$. It is straightforward to generalize
the subsequent considerations to arbitrary $n$-point correlators.
In addition, we consider a 
short-distance function $f(x)$, calculable in perturbation theory, that probes
the vacuum at distance $x\sim\Delta_f\sim Q^{-1}$. Since the function
$f(x)$ describes short-distance physics at the scale $Q\gg\Lambda$ it
is {\sl strongly peaked} compared to $g(x)$ (see Fig.\,\ref{figfandg}).
For demonstration purposes it suffices to consider an effectively
one-dimensional problem in which the chain of local power corrections 
to the purely perturbative result reads
\begin{eqnarray}
\label{mult}
\int_{-\infty}^\infty dx\,f(x)\,g(x) 
& = & 
\int dx\,dy\,f(y)\,\delta(x-y)\,g(x)
\nonumber \\[2mm] 
& = &
\int dx\,dy\,f(y)\,
\bigg[\,\sum_{n=0}^{\infty}\frac{(-1)^n}{n!}\,y^n
  \delta^{(n)}(x)\,\bigg]\,
g(x)
\nonumber\\[2mm]  & = &
\sum_{n=0}^{\infty}\bigg[\,\int dy\,f(y)\,y^n\,\bigg]\,
\left[\int dx\,\frac{(-1)^n}{n!}\delta^{(n)}(x)\,g(x)\right]
\nonumber\\[2mm]  
& = &
\sum_{n=0}^{\infty}
\,\bigg[\left.\,\bigg(i\frac{d}{dk}\bigg)^n\,\tilde
  f(k)\right|_{k=0}\,\bigg]\, 
\bigg[\,
\int\frac{dk}{2\pi}\,\frac{(ik)^n}{n!}\,\tilde g(k)
\,\bigg]   
\,.\quad
\end{eqnarray}
The functions $\tilde f$ and $\tilde g$ are the Fourier
transforms of $f$ and $g$, respectively.
From Eq.\,(\ref{mult}) it follows that 
\begin{eqnarray}
\label{multexp}
f(x) & = &
\sum_{n=0}^{\infty}
\,\bigg[\, \int dy\,f(y)\,y^n  \,\bigg]\,
\frac{(-1)^n}{n!}\,\delta^{(n)}(x)
\,,
\end{eqnarray}
which is an expansion into degenerated multipoles. 
In electrodynamics
the 3-dimensional generalization of Eq.\,(\ref{multexp}) is used to
determine the potential of a localized and static 
charge distribution $\rho(x)$ 
at distances much larger than its spatial extension (see
e.g. Ref.\,\cite{Morse1}).  
The first term on the RHS of
Eq.\,(\ref{multexp}) corresponds to the total charge times the delta
function located at the center of the charge distribution, the second
term corresponds to the dipole moment times the derivative of the
delta function, etc..
In the chain of local power corrections the derivatives of
$\delta$-functions generate higher-dimensional operators that are
obtained by covariant differentiations of the correlator
$g(x)$ at the origin $x=0$. We note that 
without specifying a regularization scheme the terms 
in the series of Eq.\,(\ref{mult}) do only exist, 
if the functions $f$ and $g$ fall off sufficiently fast in
the infrared and the ultraviolet, respectively. For $g$ this statement 
translates to the property that the local operators, which are contained 
in the corresponding chain, do not have anomalous dimensions. The
presence  of anomalous dimensions implies a regularization  
scheme, which we will ignore in this introduction and in
Secs.~\ref{sectionDOE}--\ref{sectiongluoncondensate}.

Briefly switching to 4 dimensions, we now give an example for 
gauge invariant two-point correlations leading to the function $g(x)$. The 
field strength correlator~\cite{Dosch1}
\begin{equation}
\label{gfc}
g_{\mu\nu\kappa\lambda}(x) \, \equiv \,
\mbox{Tr}\,\la g^2 \,G_{\mu\nu}(x)\,S(x,0)
\,G_{\kappa\lambda}(0)\,S(0,x)\,\ra 
\,,
\end{equation}
where
\begin{equation}
S(x,y) 
\, = \,
{\cal P}\,\e^{ig\int_0^x dz_\mu A_\mu(z)}
\,,
\label{gaugestring}
\end{equation}
generates the following chain of vacuum expectation values (VEV's)
involving local operators of increasing dimension\footnote{
The local
operators with covariant derivatives are often expressed in terms of
undifferentiated operators by appealing to the equations of motion and
Biancchi identities.}  
\begin{eqnarray}
\label{gluoncondensate}
&&
\mbox{Tr}\,\la g^2 \,G_{\mu\nu}(0)\,G_{\kappa\lambda}(0)\ra,\ 
\\[2mm]
\label{dloc}
&&
\mbox{Tr}\,\la g^2 \,G_{\mu\nu}(0)\,D_\rho\,G_{\kappa\lambda}(0)\ra,\ 
\mbox{Tr}\,\la g^2 \,G_{\mu\nu}(0)\,D_\tau D_\rho\,G_{\kappa\lambda}(0)\ra,\ 
\ldots\ ,
\end{eqnarray}
This can most easily be seen in Schwinger's fixed point gauge $x_\mu
A_\mu(x)=0$~\cite{Schwinger1} if one assumes that the integration path
in Eq.\,(\ref{gaugestring}) is a straight line.
In the limit, where
$R\equiv\Delta g/\Delta f\to\infty$, 
the lowest power correction dominates, and 
operators such as in Eq.\,(\ref{dloc}) are irrelevant in the OPE. 
The correlator of Eq.\,(\ref{gfc}) depends on the closed 
path containing the points $0$ and $x$. For the purpose of 
lattice evaluations a straight line is always assumed. 
This may be adequate for the extraction of the 
highest mass nonperturbative scale being present in the corresponding
gauge invariant correlations.   

The first VEV in Eq.\,(\ref{gluoncondensate}) is called the gluon
condensate and VEV's with odd numbers of covariant derivatives do not
contribute due to the experimentally well justified parity and
time-reversal invariance of the QCD Lagrangian.
Perturbatively, one can define the 
gluon condensate in such a way that its anomalous dimension vanishes
to all orders in $\alpha_s$.\,\cite{LoopRussians} For the channels,
which we will consider later, the anomalous dimension of the dimension
six condensates are small.\,\cite{SVZ} We will therefore ignore the
problem of anomalous dimensions in these applications. We will come
back to this issue in Sec.~\ref{sectionanomdim}.  

Let us carry on assuming a one-dimensional world. Retaining only the 
lowest power correction at $R<\infty$ corresponds to a 
saddle point approximation of the $x$-integral in (\ref{mult}). 
Expressed somewhat loosely in mathematical terms, the
strongly peaked function $f$ and the slowly varying function $g$ lie in 
a dual space defined with respect to the bilinear form  
\begin{equation}
\label{bilinear1}
(f,g) \, \equiv \, \int dx\, f(x) \,g(x)\,. 
\end{equation}
Formally, the set of functions 
\begin{equation}
e_n(x) \, \equiv \, \frac{(-1)^n}{n!}\,\delta^{(n)}(x)
\qquad\mbox{and}\qquad
\tilde e_n(x) \, \equiv \, x^n
\,,
\label{deltabasis}
\end{equation}
which are used to construct the multipole expansion in
Eq.\,(\ref{multexp}), span a basis of the dual space, with the 
orthonormality relation   
\begin{equation}
\label{ortho1}
\left(\,e_n\,,\,\tilde e_m\,\right)
\, = \, 
\delta_{nm}
\,.
\end{equation}
The series of corrections to the saddle-point
approximation, which involve the covariant derivatives of $g(x)$ at
$x=0$, is in general only asymptotic, and it is intuitively clear that
the convergence properties worsen for $R$ approaching unity.  
Much better convergence properties may be achieved by utilizing a 
multipole expansion of $f(x)$ which starts out with a saddle point
approximation based on an function of width $\Delta_f\approx Q^{-1}$
instead  of the infinitely narrow $\delta$-function. In this paper  
we present an explicit construction of such an expansion and apply it
to determine a modified version of the local OPE in
Eq.\,(\ref{Wilson}). We refer to this expansion as ``delocalized
operator expansion'' (DOE). 

The observation that the local expansion from Eq.\,(\ref{Wilson})
breaks down for $R$ approaching unity and that the corrections to the
saddle-point approximation can be substantial is well known and has
led in the past to a number of phenomenological studies considering 
directly the non-local form in the first line of Eq.\,(\ref{mult}) 
without any expansion.~\cite{Gromes1,Mikhailov1,Radyushkin1,Ball1} In
these studies the importance of 
including non-local effects has been demonstrated in a number of
cases. In this approach the short-distance function $f(x)$ has to be
computed for arbitrary values of $x$ and a specific ansatz for the
form of the non-perturbative correlator $g(x)$ has to be adopted. This
leads to the unpleasant features that the determination of the
perturbative short-distance function is considerably more complicated
than for the Wilson coefficients of the local expansion, particularly
a higher loop level, and that it necessarily involves some
model-dependence which makes the estimates of theoretical errors
difficult. The DOE has been constructed with the aim to provide an
alternative formalism for non-local effects. The DOE can be applied
with a definite power counting like the local OPE and keeps the 
computational advantages of the local expansion in the determination
of the short-distance coefficients. The new feature of the DOE is
that it relies on a generalization of the multipole expansion of the
short-distance function $f(x)$ based on functions having a width that
can be adjusted to the short-distance scale $Q$. This leads to an
improved convergence of the expansion for the 
price of having to introduce an additional parameter $\Omega$ which we call
``resolution scale''. The condensates in the DOE depend on $\Omega$ and this 
dependence accounts for non-local effects. The DOE can be used to extract
model-independent information on the non-local structure of the
non-perturbative QCD vacuum with a well defined power counting
(Sec.\,\ref{sectiongluoncondensate}), but 
also to simplify the determination of numerical predictions in a given
model for the form of non-perturbative correlation functions $g(x)$
(Secs.\,\ref{sectionquarkoniumlevel} and \ref{sectionanomdim}).

The outline of the paper is as follows. In Sec.\,\ref{sectionDOE} we
introduce the framework of the DOE 
assuming that the factorization of long- and short-distance
fluctuations can be carried out without ambiguities. The delocalized
expansion is associated with a resolution parameter $\Omega$ and, for
$\Omega\sim Q$, can lead to a series that has better convergence
properties than the local OPE. Matrix elements in the DOE
systematically sum an infinite series of local operators at each order
in $\Lambda/Q$. At lowest order in $\Lambda/Q$ the DOE implies an
$\Omega$-dependent running gluon condensate. In
Sec.\,\ref{sectionlattice} we review what is known  about the gauge
invariant gluonic field strength correlator from lattice
simulations. In Sec.\,\ref{sectionquarkoniumlevel} the nonperturbative
corrections to the  ground state energy of a heavy quarkonium system
are calculated in the local and the delocalized expansion in
$\Lambda/mv^2$. Using a lattice-inspired model for the gluon field
strength correlator, we show that the DOE has better asymptotic
convergence than the local OPE. Section\,\ref{sectiongluoncondensate}
contains an extraction of the gluon condensate from  experimental data
using charmonium sum rules and the Adler function in the $V+A$ channel
of light quark pair production. We find evidence that the gluon
condensate is indeed running. We show that our results are consistent
with the lattice measurements of the gluonic field strength
correlator. In Sec.\,\ref{sectionanomdim} we show how the DOE is
constructed for the more general case that the factorization of long-
and short-distance fluctuations is ambiguous and needs to be carried
out in a regularization scheme. We then apply the DOE to derive a
scaling relation for the ratio of meson decay constants $f_D/f_B$ at
leading order in the heavy mass expansion. A summary and an outlook
are given in Sec.\,\ref{sectionsummary}. Finally, in
App.\,\ref{appendixtensor} we show that the gluon field strength
correlator has approximately the same tensor structure as the local
gluon condensate, which simplifies computations in the investigations
in the main body of the paper. Appendix~\ref{appendixshort} contains
analytical formulae for the local short-distance coefficients needed
for the examination of quarkonium energy levels in
Sec.\,\ref{sectionquarkoniumlevel}.

\vspace{0.5cm}

\section{Delocalized operator expansion:
framework, \\evolutions equations and power counting}
\label{sectionDOE}

To construct a modified version of the multipole expansion used in
Eq.\,(\ref{multexp}) we need to find a generalization of the basis 
functions in Eq.\,(\ref{deltabasis}) satisfying 
an orthonormality relation in analogy to Eq.\,(\ref{ortho1}). 
In Ref.\,\cite{Hofmann} a lowest-order analysis based on a 4-dimensional 
spherical well was performed. Here, we construct a basis of 
the dual space by appealing to the 
orthonormality properties of the Hermite polynomials and their duals. 
For a treatment in Cartesian coordinates the following basis is well
suited: 
\begin{eqnarray}
e_n^\Omega(x) & \equiv &
\frac{\Omega^{n+1}}{\sqrt{\pi}\,n!}\,H_n(\Omega x)\,\e^{-\Omega^2 x^2}
\,,
\nonumber 
\\[2mm]
\tilde e_n^\Omega(x) & \equiv & \frac{H_n(\Omega x)}{(2 \, \Omega)^n}
\,.
\label{hermitebasis}
\end{eqnarray}
For even values of $n$ the first three Hermite polynomials read
\begin{equation}
H_0(x) \, = \, 1
\,,\qquad
H_2(x) \, = \, 4 x^2-2
\,,\qquad
H_4(x) \, = \, 16 x^4-48 x^2 + 12
\,.
\end{equation}
The width of the basis functions $e_n^\Omega$ is of order
$\Omega^{-1}$, and one recovers the basis of the local
expansion in the limit $\Omega\to\infty$:
\begin{equation}
e_n^\Omega \, \stackrel{\Omega\to\infty}{\longrightarrow} \,e_n
\qquad\mbox{and}\qquad
\tilde e_n^\Omega \, \stackrel{\Omega\to\infty}{\longrightarrow}
\,\tilde e_n
\,.
\end{equation}
An expansion of $f(x)g(x)$ using the resolution dependent 
basis $\{e_n^\Omega,\tilde e_n^\Omega\}$ generates the 
following representation of the $x$-integral in Eq.\,(\ref{mult}) 
\begin{eqnarray}
\int_{-\infty}^\infty dx\,f(x)\,g(x) 
& = &\sum_{n=0}^{\infty} f_n(\Omega) \, g_n(\Omega)
\,,
\label{multDOE}
\end{eqnarray} 
where
\begin{eqnarray}
f_n(\Omega) \, \equiv \,
(\,f\,,\,\tilde e^\Omega_n\,) & = &
\int dx \, f(x) \, \frac{H_n(\Omega x)}{(2 \, \Omega)^n}
\nonumber\\[2mm]
& = &
\left.
\frac{1}{(2 \, \Omega)^n}\,
H_n\bigg(\Omega\bigg(i\frac{d}{dk}\bigg)\bigg) 
\,\tilde f(k) \right|_{k=0}
\,,
\label{fdefconvolution}
\\[3mm]
g_n(\Omega) \, \equiv \,
(\,e^\Omega_n\,,\,g\,) & = &
\int
dx\, \frac{\Omega^{n+1}}{\sqrt{\pi}\,n!}\,H_n(\Omega x)
\,\e^{-\Omega^2 x^2} \,g(x)
\nonumber\\[2mm]
& = &
\int\frac{dk}{2\pi}\,\frac{(i k)^n}{n!}\,
\e^{-\frac{k^2}{4\Omega^2}} \,\tilde g(k)
\,. 
\label{gndefconvolution}
\end{eqnarray}
We call $\Omega$ the ``resolution parameter''. It is intuitively 
clear that the series in Eq.\,(\ref{multDOE}) has better convergence
properties, if $\Omega$ is of order $Q$, because in this case the term
$(f,\tilde e_0^{\Omega})e_0^\Omega(x)$ accounts much better for the form
of $f(x)$ than for $\Omega=\infty$.  The terms $f_n(\Omega)$ are the
resolution-dependent short-distance coefficients, and the
$g_n(\Omega)$ are the resolution-dependent condensates.
Note that, if $f(x)$ is positive definite within its main support 
$|x|\,<\,Q^{-1}$, then $f_n(\infty)>0$ 
for $n$ not too large. On the other hand, if $g(x)$ is positive
definite and monotonic decreasing for $x>0$, then
$g_n(\Omega)<g_n(\Omega^\prime)$ for $\Omega<\Omega^\prime$. 

The relation between basis
functions for different resolution parameters $\Omega$ and
$\Omega^\prime$ can be obtained from the properties of the Hermite
polynomials and reads
\begin{eqnarray}
\tilde e_n^\Omega(x) & = &
\sum_{m=0}^{\infty} a_{nm}(\Omega,\Omega^\prime)\,
\tilde e_m^{\Omega^\prime}(x)
\nonumber\\[2mm]
e_n^\Omega(x) & = &
\sum_{m=0}^{\infty} a_{mn}(\Omega^\prime,\Omega)\,
e_m^{\Omega^\prime}(x)
\,,
\label{changeofbasis}
\end{eqnarray}
where
\begin{equation}
a_{nm}(\Omega,\Omega^\prime) \, = \,
\left\{
\begin{array}{ll}
\frac{\displaystyle n!}{{\displaystyle m!}\,\big(\mbox{$\frac{n-m}{2}$}\big)!}\,
\bigg({\displaystyle \frac{\Omega^2-\Omega^{\prime\,2}}
      {4\,\Omega^2\,\Omega^{\prime\,2}}}\bigg)^{\big(\mbox{$\frac{n-m}{2}$}\big)}
\,, & n-m \ge 0\,\,\mbox{and even}\,,
\\[6mm]
0\,, & \mbox{otherwise}\,.
\end{array}
\right.
\label{anmdef}
\end{equation}
Equation\,(\ref{anmdef}) leads to the following relations between the
short-distance coefficients $f_n$ and the corresponding condensates
$g_n$ for different resolutions $\Omega$ and $\Omega^\prime$: 
\begin{eqnarray}
f_n(\Omega) & = &
\sum_{m=0}^{\infty} a_{nm}(\Omega,\Omega^\prime)\,
f_m(\Omega^\prime)
\label{frelations}
\\[2mm]
g_n(\Omega) & = &
\sum_{m=0}^{\infty} a_{mn}(\Omega^\prime,\Omega)\,
g_m(\Omega^\prime)
\,.
\label{fgrelations}
\end{eqnarray}
The coefficients $a_{nm}(\Omega,\Omega^\prime)$ satisfy the
composition property 
\begin{equation}
\label{com}
\sum_{i=0}^\infty
a_{ni}(\Omega,\Omega^\prime)\,
a_{im}(\Omega^\prime,\Omega^{\prime\prime})
\, =  \,
a_{nm}(\Omega,\Omega^{\prime\prime})\,,
\end{equation}
and for equal resolution we have
\begin{equation}
\label{inverse}
a_{nm}(\Omega,\Omega)
\, = \, \delta_{nm}
\,.
\end{equation}
Obviously, the transformations $a_{nm}(\Omega,\Omega^\prime)$ form a
group. 
It is instructive to specify the relations in 
Eqs.\,(\ref{frelations}) and (\ref{fgrelations}) for
$\Omega^\prime=\infty$, 
\begin{eqnarray}
f_0(\Omega) & = & f_0(\infty)\,, 
\nonumber\\[2mm]
f_2(\Omega) & = & f_2(\infty)\, - \, \frac{1}{2\,\Omega^2} f_0(\infty)\,,
\nonumber\\[2mm]
f_4(\Omega) & = & f_4(\infty)\, - \, \frac{3}{\Omega^2} f_2(\infty) 
              \, + \, \frac{3}{4\,\Omega^4}\,f_0(\infty)\,, 
\nonumber\\[2mm] 
	    &\vdots& 
\nonumber\\[2mm] 
f_n(\Omega) & = & \sum_{i=0}^{[n/2]}\frac{n!}{(n-2i)!\,i!}
	    \left(-\frac{1}{4\Omega^2}\right)^i\,f_{n-2i}(\infty)\,,
\nonumber\\[2mm]
g_n(\Omega) & = & 
\sum_{i=n}^\infty \, \frac{(n+2i)!}{n!\,i!}\,
\Big(\frac{1}{4\,\Omega^2}\Big)^i\,g_{n+2i}(\infty)
\,,
\label{fgrelationOmegainfty}
\end{eqnarray}
where the explicit expressions for $f_n(\infty)$ and $g_n(\infty)$ can 
be read off Eq.\,(\ref{mult}). We see that each short-distance
coefficient  $f_n(\Omega)$ can be expressed in terms of a {\it finite}
linear combination of the local Wilson coefficients $f_i(\infty)$ for
$i\le n$. In particular, the short-distance coefficient $f_0(\Omega)$
and the Wilson coefficient of the first term in the local expansion
coincide since $H_0(\Omega x)=1$, i.e. the short-distance coefficient
in the saddle-point approximation is $\Omega$-independent.  
On the other hand, the resolution-dependent condensates $g_n(\Omega)$  
are related to an {\it infinite} sum of local condensates with
additional covariant derivatives multiplied by $\Omega$-dependent
coefficients. From now on we adopt the 
language saying that resolution-dependent condensates 
of a given dimension can be expressed as an infinite sum of local 
condensates of equal and higher dimension. And similarly we say that
each  resolution-dependent short-distance coefficient associated with a
condensate of a certain dimension can be expressed 
as a finite sum of local short-distance coefficients associated with 
condensates of equal or smaller dimensions.

Note that in Eq.\,(\ref{multDOE}) the term
$f_0(\infty)g_2(\infty)/(2\Omega^2)$ contained in
$f_0(\Omega)g_0(\Omega)$ is cancelled by the term 
$-f_0(\infty)g_2(\infty)/(2\Omega^2)$ in
$f_2(\Omega)g_2(\Omega)$ which exemplarily shows the 
reshuffling of the OPE. As we have argued above, 
the Wilson coefficients $f_n(\infty)$ have equal sign 
for $n$ not too large and therefore the 
short-distance coefficient $f_2(\Omega)$ can be much smaller than 
$f_2(\infty)$ for $\Omega\sim Q$. This feature and the fact that 
$g_n(\Omega)<g_n(\infty)$ under the conditions mentioned above
can, at least for small $n$, severely suppress higher order
contributions. The DOE of the current correlator in
Eq.\,(\ref{Wilson}) has the same parametric counting in powers of
$\Lambda/Q$ as the OPE, as long as
$\Omega$ is not chosen parametrically smaller than $Q$. Taking into account that
$g_n\sim\Lambda^n$ and $f_n\sim Q^{-n}$ in the local expansion, we also 
have 
\begin{equation}
g_n(\Omega)\sim \Lambda^n \sum_{i}\Big(\frac{\Lambda}{\Omega}\Big)^i \sim
\Lambda^n
\,, \qquad
f_n(\Omega)\sim Q^{-n} \sum_{i}\Big(\frac{Q}{\Omega}\Big)^i \sim Q^{-n}
\quad\mbox{}
\end{equation}
in the DOE as long as $\Omega\gg Q$. 
At finite resolution, however, one obtains additional summations
of powers of $Q/\Omega$ and $\Lambda/\Omega$ in the short-distance
coefficients and the condensates, respectively. Adopting a
lattice-inspired model, we will see in   
Sec.~\ref{sectionquarkoniumlevel} that for $\Omega\sim Q$ the
expansion in powers of $\Lambda/Q$ has an additional suppression by
powers of a small number as compared to the expansion for $\Omega=\infty$.

It is straightforward to extend the previous results 
to an arbitrary number of dimensions by applying the
resolution-dependent expansion independently to each
coordinate. This also accounts for the treatment of arbitrary $n$-point 
correlations in an arbitrary number of dimensions.
For convenience we consider here the same 
resolution parameter $\Omega$ for each coordinate. It  
is straightforward to derive the resolution-dependent expansion for
a more general situation. 

If $f=f(x_1,\ldots, x_d)$ is a strongly peaked 
short-distance function in $d$ dimensional Euclidean space-time 
the analogous expansion to Eq.\,(\ref{multDOE}) reads
\begin{eqnarray}
\int_{-\infty}^\infty d^dx\,f(x_1,\ldots, x_d)\,g(x_1,\ldots, x_d) 
& = & 
\sum_{n_1,\ldots,n_d=0}^{\infty} f_{n_1,\ldots,n_d}(\Omega) \, 
g_{n_1,\ldots,n_d}(\Omega)
\,,\quad
\label{multDOEddim}
\end{eqnarray} 
where
\begin{eqnarray}
f_{n_1,\ldots,n_d}(\Omega) & \equiv & 
\int d^dx \, f(x_1,\ldots, x_d) \, 
\bigg[\,
\prod_{i=1}^d \frac{H_{n_i}(\Omega x_i)}{(2 \, \Omega)^{n_i}}
\,\bigg]
\nonumber\\[2mm]
& = &
\left.
\bigg[\,
\prod_{i=1}^d \, \frac{1}{(2 \, \Omega)^{n_i}}\,
H_{n_i}\bigg(\Omega\bigg(i\frac{d}{dk_i}\bigg)\bigg) 
\,\bigg]
\,\tilde f(k_i,\ldots ,k_d) \right|_{k_1,\ldots ,k_d=0}
\,,
\label{fmultidim}
\\[2mm]
g_{n_1,\ldots,n_d}(\Omega) & \equiv & 
\int
d^dx\, 
\bigg[\,
\prod_{i=1}^d \frac{\Omega^{n_i+1}}{\sqrt{\pi}\,n_i!}\,H_{n_i}(\Omega x_i)
\,\bigg]
 \,\e^{-\Omega^2 x^2}
\,g(x_1,\ldots, x_d)
\nonumber\\[2mm]
& = &
\int\frac{d^dk}{(2\pi)^d}\,
\bigg[\,
\prod_{i=1}^d \,\frac{(i k_{n_i})^{n_i}}{n_i!}
\,\bigg]
\,\e^{-\frac{k^2}{4\Omega^2}} \,\tilde g(k_1,\ldots ,k_d)
\,. 
\label{fandgddim}
\end{eqnarray}
The relation between the short-distance coefficients and the
condensates in Eqs.\,(\ref{fmultidim}) and (\ref{fandgddim}) for
different resolution 
parameters is derived from Eqs.\,(\ref{changeofbasis}) for each index
$n_i$ in analogy to the expressions in Eqs.\,(\ref{frelations}) and
(\ref{fgrelations}). 
As in the one-dimensional case each short-distance coefficient
$f_{n_1,\ldots,n_d}(\Omega)$ can be 
expressed in terms of a {\it finite} linear combination of the
local Wilson coefficients $f_{i_1\ldots,i_d}(\infty)$ for 
$i_k\le n_k$, and the short-distance coefficient 
$f_{0,\ldots,0}(\Omega)$ is resolution-independent.
The resolution-dependent condensates $g_{n_1,\ldots,n_d}(\Omega)$ 
are related to an {\it infinite} sum of local condensates 
$g_{i_1,\ldots,i_d}(\infty)$ for $i_k\ge n_k$. 
For example, assuming that  $g$ is a function of the distance
$|x|=(x_1^2+\ldots +x_d^2)^{1/2}$ only, one finds
\begin{eqnarray}
g_{0,\ldots,0}(\Omega) & = & 
g_{0,\ldots,0}(\infty) + 
\frac{1}{2Q^2}\,\Big(\,
  g_{2,\ldots,0}(\infty)+ \ldots +g_{0,\ldots,2}(\infty)\,\Big)
+\ldots
\nonumber\\[2mm] 
& = & g(0)+\frac{1}{4Q^2}\pd_\rho\pd_\rho g(0) + \ldots 
\,.
\label{g0000}
\end{eqnarray}

It is an easy exercise to derive evolution equations for the
resolution-dependent short-distance coefficients and the
condensates. In the one-dimensional case the evolution equations are
obtained directly from Eqs.\,(\ref{frelations}) and
(\ref{fgrelations}) and read
\begin{eqnarray}
\frac{d}{d\,\Omega}\,f_n(\Omega) & = &
\frac{(n-1)\,n}{2\,\Omega^3}\,f_{n-2}(\Omega)
\,,
\nonumber\\[2mm]
\frac{d}{d\,\Omega}\,g_n(\Omega) & = &
-\,\frac{(n+1)(n+2)}{2\,\Omega^3}
\,g_{n+2}(\Omega)
\,.
\label{fandgevolution}
\end{eqnarray} 
Equations~(\ref{fandgevolution}) imply that there is only mixing 
between terms that differ by two mass dimensions. This is a
consequence of the specific choice of the basis functions in
Eqs.\,(\ref{hermitebasis}), which are constructed from the Hermite
polynomials and the Gaussian function.
For the short-distance coefficients lower dimensional terms always mix into
higher dimensional ones. Prescribing short-distance coefficients at
$\Omega=\infty$, the solutions are given by
Eqs.\,(\ref{fgrelationOmegainfty}). For the condensates  
higher dimensional terms always mix into lower dimensional ones. Their
evolution equations are much more interesting because they show that
knowing the leading condensate $g_0(\Omega)$ implies that all
subleading condensates of higher dimensions are obtained by
differentiations with respect to $\Omega$. The analogous features also
exist for the $d$-dimensional generalization of the evolution
equations, which have the form
\begin{eqnarray}
\frac{d}{d\,\Omega}\,f_{n_1,\ldots,n_d}(\Omega) & = &
\frac{(n_1-1)\,n_1}{2\,\Omega^3}\,f_{n_1-2,\ldots,n_d}(\Omega)
+ \ldots +
\frac{(n_d-1)\,n_d}{2\,\Omega^3}\,f_{n_1,\ldots,n_d-2}(\Omega)
\,,
\nonumber\\[2mm]
\frac{d}{d\,\Omega}\,g_{n_1,\ldots,n_d}(\Omega) & = &
-\,\frac{(n_1+1)(n_1+2)}{2\,\Omega^3}
\,g_{n_1+2,\ldots,n_d}(\Omega)
- 
\nonumber\\ & & \hspace{1cm}
\ldots -
\frac{(n_d+1)(n_d+2)}{2\,\Omega^3}
\,g_{n_1,\ldots,n_d+2}(\Omega)
\,.
\label{fandgevolutionddim}
\end{eqnarray} 

\vspace{0.5cm}

\section{The gluonic field strength correlator} 
\label{sectionlattice}

Since it is the most important 2-point function, which hence we will 
heavily draw upon in subsequent sections, we review here what is known
from lattice calculations about the gauge invariant, gluonic field
strength correlator~(\ref{gfc}). Without constraining generality 
the following parametrization of this function in Euclidean space-time 
was introduced in Ref.\,\cite{Dosch1}: 
\begin{eqnarray}
g_{\mu\nu\kappa\lambda}(x) & = & 
(\,\delta_{\mu\kappa}\delta_{\nu\lambda}-\delta_{\mu\lambda}\delta_{\nu\kappa}\,)
\,\Big[\,D(x^2)+D_1(x^2)\,\Big]
\nonumber\\[2mm]  & &
+\,(\,x_\mu x_\kappa\delta_{\nu\lambda}-x_\mu x_\lambda\delta_{\nu\kappa}+
x_\nu x_\lambda\delta_{\mu\kappa}-x_\nu x_\kappa\delta_{\mu\lambda}\,)
\,\frac{\partial D_1(x^2)}{\partial x^2}
\,.
\label{pargfc}
\end{eqnarray}
To separate perturbative from nonperturbative contributions the scalar
functions $D$ and $D_1$ are usually fitted as
\begin{eqnarray}
D(x^2) & = & 
A_0\,\exp\bigg(\!-\frac{|x|}{\lambda_A}\bigg)
+\frac{a_0}{|x|^4}\,\exp\bigg(\!-\frac{|x|}{\lambda_a}\bigg)\,,
\nonumber\\[2mm] 
D_1(x^2) & = & 
A_1\,\exp\bigg(\!-\frac{|x|}{\lambda_A}\bigg)+
\frac{a_1}{|x|^4}\,\exp\bigg(\!-\frac{|x|}{\lambda_a}\bigg)
\,. 
\label{gfcfit}
\end{eqnarray}
The power-like behavior in Eqs.\,(\ref{gfcfit}) at small $|x|$ is
believed to catch most of the perturbative physics, although it is
known that partially summed perturbation theory may generate
ultraviolet-finite 
contributions for $|x|=0$ as well. However, magnitude estimates are
difficult because the integration over the corresponding  
poles in the Borel plane is known to be ambiguous.\,\cite{AMueller} 
We ignore this subtlety and assume that the 
purely exponential terms in Eqs.\,(\ref{gfcfit}) exclusively carry
nonperturbative information. In a recent unquenched  
lattice simulation~\cite{Delia} (see also Ref.\,\cite{Brambilla1})
where the gluon field strength correlator was measured with a
resolution of 
$a^{-1}\approx (0.1\,\mbox{fm})^{-1} \approx 2$\,GeV
between $3$ and $8$ lattice spacings it was found that  
\begin{equation}
\label{lat}
\frac{A_0}{A_1} \, \approx \, 9
\qquad\mbox{and}\qquad
\lambda_A^{-1} \, \approx \, 0.7\,\mbox{GeV} 
\,.
\end{equation}
It is conspicuous that the inverse correlation length is somewhat
larger than the typical hadronization scale $\Lambda_{\rm QCD}$. 
Whereas the actual size of $A_0$ and $A_1$ depend quite strongly on
whether quenched or unquenched simulations are carried out and which
values for the light quark masses were assumed, the ratio $A_0/A_1$
and the correlation length $\lambda_A$ were found to be quite
stable.\,\cite{Delia}

Because $A_0\gg A_1$ we may neglect the $D_1$ contribution in
Eq.\,(\ref{pargfc}) at the level of precision intended in our
subsequent analyses and write 
\begin{equation}
g_{\mu\nu\kappa\lambda}^{\rm non-pert}(x) \, = \, 
\frac{1}{12}\, 
(\,\delta_{\mu\kappa}\delta_{\nu\lambda}-\delta_{\mu\lambda}\delta_{\nu\kappa}\,)
\,g(|x|)
\,.
\label{pargfcsimple}
\end{equation}
This simplifies the computations involved in the DOE since in this
approximation the gluon field strength correlator has the same
$x$-independent tensor structure as the local gluon condensate 
\begin{equation}
g_{\mu\nu\kappa\lambda}^{\rm non-pert}(0) \, = \,
(\,\delta_{\mu\kappa}\delta_{\nu\lambda}-\delta_{\mu\lambda}\delta_{\nu\kappa}\,)
\,\frac{\pi^2}{6}\,
\Big\langle 0\Big|
\,\frac{\alpha_s}{\pi}\, G^a_{\rho\sigma}(0)G^a_{\rho\sigma}(0)
\,\Big|0\Big\rangle
\,.
\end{equation}
It is shown in App.\,\ref{appendixtensor} that the lattice-implied
dominance of the  tensor structure in Eq.\,(\ref{pargfcsimple}) is
consistent with a  determination of  
\begin{equation}
\mbox{Tr}\,\Big\langle\, 
g^2 G_{\mu\nu}(0)\,D_{\alpha}D_{\beta}\,
G_{\kappa\lambda}(0)\,\Big\rangle 
\label{D2G2}
\end{equation}
using equations of motion, Biancchi identities, and phenomenologically
obtained values for the VEV's  $\la g^2 j_\mu^a j_\mu^a\ra$ and 
$\la g f_{abc} G^a_{\mu\nu}G^b_{\nu\lambda}G^c_{\lambda\mu}\ra$. 

We note that due to the cusp of the exponential ansatz in
Eq.\,(\ref{gfcfit}), higher derivatives at $x=0$ have nonlogarithmic
UV-singularities that can only be defined in a regularization
scheme. So assuming an  exponential behavior down to arbitrary small
distances would cause inconsistencies in the Wilson OPE for operators
containing covariant derivatives, which are known to have logarithmic
anomalous dimensions. Since unquenched lattice simulations
have not been carried out for distances below $3$ lattice spacings,
the simple exponential ansatz is invalid for some region around $x=0$
with $|x|<\lambda_A$. 

\vspace{0.5cm}

\section{Nonperturbative corrections to the heavy \\
quarkonium ground state level}
\label{sectionquarkoniumlevel}

Among the early applications of the OPE in QCD was the analysis of
non\-per\-tur\-ba\-tive effects in heavy quarkonium
systems.\,\cite{Voloshin1,Leutwyler1,Voloshin2} Heavy 
quarkonium systems are non\-re\-la\-ti\-vis\-tic quark-antiquark bound
states for which there is the following hierarchy of the relevant  
physical scales $m$ (heavy quark mass), $m v$
(relative momentum), $m v^2$ (kinetic energy) and $\Lambda_{\rm QCD}$:
\begin{equation}
\label{condition1}
m \, \gg \, m v \, \gg \, m v^2 \, \gg \, \Lambda_{\rm QCD}
\,.
\end{equation}
Thus the spatial size of the quarkonium system $\sim (m v)^{-1}$ is
much smaller than the typical dynamical time scale $\sim (m
v^2)^{-1}$. We note that in practice the last of the conditions in
Eq.\,(\ref{condition1}), which relates the vacuum correlation length
with the quarkonium energy scale, is probably not satisfied for any
known quarkonium state, not even for $\Upsilon$ mesons~\cite{Gromes1}.
Only for top-antitop quark threshold production
condition~(\ref{condition1}) is believed to be a viable
assumption.~\cite{topthreshold} 

In this section we demonstrate the DOE for the
nonperturbative corrections to the $n^{\,2s+1}L_j=1^{\,3}S_1$ ground
state for different values of $m$ in the model defined in
Eq.\,(\ref{gmodel}). We adopt the local version of the
multipole expansion (OPE) for the expansion in the ratios of the
scales $m$, $m v$ and $m v^2$. The resolution dependent expansion
(DOE) is applied with respect to the ratio of  the scales $m v^2$ and
$\Lambda_{\rm QCD}$.   
The former expansion amounts to the usual treatment of the dominant
perturbative dynamics by means of a nonrelativistic two-body
Schr\"odinger equation. The interaction with the
nonperturbative vacuum is accounted for by two insertions of the local 
$\bmx\bmE$ dipole operator, $\bmE$ being the chromoelectric
field.\,\cite{Voloshin1} The chain of VEV's of the two gluon operator
with increasing numbers of covariant derivatives 
times powers of quark-antiquark octet propagators~\cite{Voloshin1},
i.e. the expansion in  $\Lambda/m v^2$, is treated in the DOE.
Our examination is not intended to represent a phenomenological study
of non-perturbative effects in heavy quarkonium energy levels, 
but to demonstrate the result of the DOE for a natural choice of the
resolution scale in comparison to the local expansion and the exact result
in a given model. 

At leading order in the local multipole expansion with respect to the
scales $m$, $m v$, and $m v^2$ the expression for the nonperturbative
corrections to the ground state energy in a form analogous to
Eqs.\,(\ref{mult}) and (\ref{multDOE}) reads
\begin{equation}
E^{np} 
\, = \,
\int_{-\infty}^\infty\! dt \, f(t) \, g(t)
\,,
\label{Enonpertdef}
\end{equation}
where
\begin{equation}
f(t) \, =  \, \frac{1}{36}\,
\int \frac{dq_0}{2\pi}\,\e^{i q_0(it)}\,
\int\! d^3\bmx \int\! d^3\bmy\,
\phi(x)\,(\bmx\bmy)\,
G_O\bigg(\bmx,\bmy,-\frac{k^2}{m}-q_0\bigg)\,\phi(y)
\,,
\label{Enonpertfdef}
\end{equation}
with
\begin{eqnarray}
G_O\bigg(\bmx,\bmy,-\frac{k^2}{m}\bigg)
& = &
\sum_{l=0}^\infty\,(2l+1)\,P_l\bigg(\frac{\bmx\bmy}{x y}\bigg)\,
G_l\bigg(x,y,-\frac{k^2}{m}\bigg)
\,,
\nonumber\\[2mm]
G_l\bigg(x,y,-\frac{k^2}{m}\bigg) 
& = &
\frac{m k}{2\pi}\,(2kx)^l\,(2ky)^l\,\e^{-k(x+y)}\,
\sum_{s=0}^\infty \,\frac{L_s^{2l+1}(2kx)\,L_s^{2l+1}(2ky)\,s!}
          {(s+l+1-\frac{m\,\alpha_s}{12\,k})\,(s+2l+1)!}
\,,
\nonumber\\[2mm]
\phi(x) & = & \frac{k^{3/2}}{\sqrt{\pi}}\,\e^{-k x}
\,,\qquad
\nonumber\\[2mm]
k & = & \frac{2}{3}\,m\alpha_s
\end{eqnarray}
The term $G_O$ is the quark-antiquark octet
Green-function~\cite{Voloshin2},
and $\phi$ denotes the ground state wave function. The functions $P_n$ and
$L_n^k$ are the Legendre and Laguerre polynomials, respectively.
Since we neglect the
spatial extension of the quarkonium system with respect to the
interaction with the nonperturbative vacuum, the insertions of the
$\bmx\bmE$ operator probe only the temporal correlations in the
vacuum, which effectively renders the problem one-dimensional. We note
that $t$ is the Euclidean time. This is the origin of the term $it$
in the exponent appearing in the definition of the function $f$. For $f$ 
it is implied that possible odd components are removed through the
operation $f(t)\to (f(t)+f(-t))/2$ because $g$ is an even function.
\begin{figure}[t!] 
\begin{center}
\leavevmode
\epsfxsize=3.5cm
\epsffile[260 430 420 720]{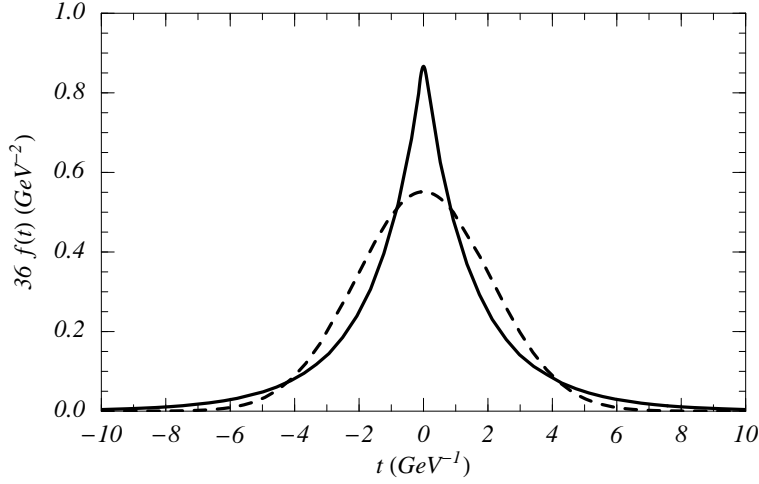}
%
%
\vspace{0cm}
 \caption{\label{figffunction} 
The perturbative short-distance function $f(t)$ for $m=5$~GeV and
$\alpha_s=0.39$ (solid line) and the leading term in the delocalized
multipole expansion of $f(t)$ for $\Omega=k^2/m$ (dashed line). 
}
 \end{center}
\end{figure}
In Fig.\,\ref{figffunction} the function $f(t)$ (solid line) is
displayed for $m=5$\,GeV and $\alpha_s=0.39$. Since the spatial
extension of the quarkonium system is neglected and the average time
between interactions with the vacuum is of the order of the inverse
kinetic energy, the characteristic width of $f$ is of order 
$(m v^2)^{-1}\sim(m\alpha_s^2)^{-1}$. As a 
comparison we have also displayed  the function 
$[\int dt^\prime f(t^\prime)\tilde e_0^\Omega(t^\prime)]
e_0^\Omega(t)$ for $\Omega=k^2/m$ (dashed line), which is the
leading term in the delocalized multipole expansion of $f$. The values
of the first few local multipole moments $f_n(\infty)$, which
correspond to the local Wilson coefficients, read
\begin{equation}
\begin{array}{ll}
f_0(\infty) \, = \, 1.6518\,\frac{m}{36\,k^4}\,, \qquad 
&  
f_2(\infty) \, = \, 1.3130\,\frac{m^3}{36\,k^8}\,, 
\\[3mm]
f_4(\infty) \, = \, 7.7570\,\frac{m^5}{36\,k^{12}}\,,  
&  
f_6(\infty) \, = \, 130.492\,\frac{m^7}{36\,k^{16}}\,, 
\\[3mm]
f_8(\infty) \, = \, 4474.1\,\frac{m^9}{36\,k^{20}}\,,
&  
f_{10}(\infty) \, = \,262709.3\,\frac{m^{11}}{36\,k^{24}}\,,\ldots\,.
\end{array}
\label{fns}
\end{equation}
The term $f_0(\infty)$
agrees with Ref.\,\cite{Leutwyler1,Voloshin2} and $f_2(\infty)$ with
Ref.\,\cite{Pineda}. The results for $f_{n>2}(\infty)$ are new.
Analytical expressions for the numbers shown in Eqs.\,(\ref{fns}) are
given in App.\,\ref{appendixshort}.

Let us compare the local expansion of $E^{np}$ with the
resolution-dependent expansion using the basis functions of 
Eq.\,(\ref{hermitebasis}). For the nonperturbative gluonic field strength
correlator we use a lattice inspired model of the form
\begin{eqnarray}
g(t) & = & 12\,A_0 \,\exp\Big(-\sqrt{t^2+\lambda_A^{2}}/\lambda_A + 1\Big)
\,,
\nonumber\\[2mm]
A_0 & = & 0.04~\mbox{GeV}^{4}\,,
\qquad
\lambda_A^{-1} \, = \, 0.7~\mbox{GeV}
\,,
\label{gmodel}
\end{eqnarray}
This model has an exponential large-time behavior according to
Eq.\,(\ref{gfcfit}) and a smooth behavior for small $t$. The 
local dimension $4$ gluon condensate in this model is
\begin{equation}
\Big\langle\frac{\alpha_s}{\pi}\,G_{\mu\nu}^a G_{\mu\nu}^a\Big\rangle
\, = \, 
\frac{6\,A_0}{\pi^2} 
\, = \,  0.024\,\,\mbox{GeV}^4
\,.
\end{equation}
We note that for the purpose of this examination the exact form of the
model for the nonperturbative gluonic field strength
correlator is not an important issue as long as the derivatives of
$g(t)$ at $t=0$ are well defined, see
e.g. Refs.\,\cite{Gromes1,Balitsky1} for different choices of models. 

In Tab.\,\ref{tabquarkonium} we have displayed the exact result in the
model~(\ref{gmodel}) and
the first four terms of the resolution-dependent expansion of $E^{np}$ 
for the quark masses $m=5,25,45,90,175$~GeV and for $\Omega=\infty$
and $\Omega=k^2/m$. For each value of the quark mass the strong coupling has
been fixed by the relation $\alpha_s = \alpha_s(k)$. 
\begin{table}[t!]  
\begin{center}
\begin{tabular}{|c|c|c|c||r|r|r|r|r|} \hline
\multicolumn{5}{|c|}{} & \multicolumn{2}{|c|}{$\Omega=\infty$} 
                  & \multicolumn{2}{|c|}{$\Omega=k^2/m$}
 \\ \hline
$m$ &  & $k^2/m$ & $E^{np}$ & & $f_ng_n$ & $\sum^n_{i=0}f_ig_i$ 
                              & $f_ng_n$ & $\sum^n_{i=0}f_ig_i$
 \\ 
(GeV) & \raisebox{1.5ex}[-1.5ex]{$\alpha_s$} & (MeV) & (MeV)
   & \raisebox{1.5ex}[-1.5ex]{$n$} & (MeV) & (MeV) & (MeV) & (MeV)
 \\ \hline\hline
$5$ &   $0.39$ & $0.338$ & $24.8$ & 
           0 & $38.6$      &  $38.6$    & $24.2$    & $24.2$
\\ \cline{5-9}
& & & &    2 & $-65.7$     & $-27.2$    & $-3.9$    & $20.3$
\\ \cline{5-9}
& & & &    4 & $832.7$     & $805.5$    & $12.1$    & $32.4$
\\ \cline{5-9}
& & & &    6 & $-35048.0$  & $-34242.4$  & $-43.1$  & $-10.8$
\\ \hline\hline
$25$ &  $0.23$ & $0.588$ & $12.6$ & 
           0 & $16.0$    &  $16.0$    &  $12.8$ &  $12.8$
\\ \cline{5-9}
& & & &    2 & $-9.0$    &  $7.0$     &  $-1.2$ &  $11.5$
\\ \cline{5-9}
& & & &    4 & $37.8$    &  $44.8$    &  $2.6$  &  $14.1$
\\ \cline{5-9}
& & & &    6 & $-526.6$  &  $-481.8$  &  $-6.7$ &  $7.4$
\\ \hline\hline
$45$ &  $0.19$ & $0.722$ & $4.9$ & 
           0 & $5.9$    &  $5.9$    &  $5.0$   &  $5.0$
\\ \cline{5-9}
& & & &    2 & $-2.2$   &  $3.7$    &  $-0.4$  &  $4.6$
\\ \cline{5-9}
& & & &    4 & $6.1$    &  $9.8$    &  $0.7$   &  $5.3$
\\ \cline{5-9}
& & & &    6 & $-56.4$  &  $-46.6$  &  $-1.4$  &  $3.8$
\\ \hline\hline
$90$ &  $0.17$ & $1.156$ & $1.05$ & 
           0 & $1.15$   &  $1.15$  &  $1.07$   &  $1.07$
\\ \cline{5-9}
& & & &    2 & $-0.17$  &  $0.98$  &  $-0.04$  &  $1.02$
\\ \cline{5-9}
& & & &    4 & $0.18$   &  $1.16$  &  $0.04$   &  $1.07$
\\ \cline{5-9}
& & & &    6 & $-0.65$  &  $0.51$  &  $-0.07$  &  $1.00$
\\ \hline\hline
$175$ & $0.15$ & $1.750$ & $0.245$ & 
           0 & $0.258$   &  $0.258$  &  $0.249$   &  $0.249$
\\ \cline{5-9}
& & & &    2 & $-0.016$  &  $0.242$  &  $-0.005$  &  $0.244$
\\ \cline{5-9}
& & & &    4 & $0.008$   &  $0.250$  &  $0.003$   &  $0.247$
\\ \cline{5-9}
& & & &    6 & $-0.012$  &  $0.237$  &  $-0.003$  &  $0.244$
\\ \hline
\end{tabular}
\caption{\label{tabquarkonium} 
Nonperturbative corrections to the heavy quarkonium ground state level
at leading order in the multipole expansion with respect to the scales
$m$, $mv$ and $mv^2$ for various quark masses $m$ based on the model
in Eq.\,(\ref{gmodel}). Displayed are the exact result and the first
few orders in the DOE for $\Omega=\infty$ and $\Omega=k^2/m$.
The numbers are rounded off to units of $0.1$, $0.01$ or $0.001$~MeV. 
}
\end{center}
\vskip 3mm
\end{table}
We note that the series are all asymptotic, i.e.\,\,they are not
convergent for any resolution. The local expansion ($\Omega=\infty$)
is quite badly behaved 
for smaller quark masses because for $k^2/m\lsim\lambda_A^{-1}$ any
local expansion is meaningless. In particular, for $m=5$~GeV the
subleading dimension $6$ term is already larger than the parametrically
leading dimension $4$ term. This is consistent with the size of the
dimension $6$ term based on a phenomenological estimate of the local
dimension $6$ condensate in Eq.\,(\ref{D2G2}), see
App.\,\ref{appendixtensor}. 
For quark masses, where $k^2/m\gsim\lambda_A^{-1}$, the local
expansion is reasonably good. However, for finite resolution
$\Omega=k^2/m$, the size of higher order terms is considerably smaller
than in the local expansion for all quark masses, and the series
appears to be much better behaved. The size of the order $n$ term is
suppressed by approximately a factor $2^{-n}$ as compared to the order
$n$ term in the local expansion. We find explicitly that terms in
the series with larger $n$ decrease more quickly for finite resolution
scale as compared to the local expansion. One also observes that even
in the case $k^2/m < \lambda_A^{-1}$, where the leading term of the
local expansion overestimates the exact result, the leading
term in the delocalized expansion for $\Omega=k^2/m$ agrees with the 
exact result within a few percent. It is intuitively clear that this
feature is a general property of the delocalized expansion, and we
believe that it should apply to any problem for which the local expansion
in the ratio of two scales breaks down because the ratio is not
sufficiently small. 

We would like to emphasize again that the previous analysis is not
intended to provide a phenomenological determination of
nonperturbative 
corrections to the heavy quarkonium ground state energy level, but
rather to demonstrate the DOE within a specific model. For a realistic 
treatment of the nonperturbative contributions in the heavy quarkonium
spectrum a model-independent analysis should be carried out. In
addition, also higher orders in the local multipole expansion with
respect to the  ratios of scales $m$, $m v$ and $m v^2$ should be
taken into account, which have been neglected here. These corrections
might be substantial, in particular for smaller quark masses. However,
having in mind an application to the bottomonium spectrum, we believe
that our results for the expansion  in $\Lambda/mv^2$ indicate that
going  beyond the leading term in the OPE for the bottomonium ground
state is probably meaningless and that calculations based on the DOE
with a suitable choice of resolution should be more reliable.

\vspace{0.5cm}

\section{Extraction of the running gluon condensate}
\label{sectiongluoncondensate}

The resolution-dependent condensates are either determined
phenomenologically from experimental data or from lattice
measurements. It is the purpose of this section to extract the
resolution-dependent dimension $4$ condensate from an ana\-ly\-sis of   
moment ratios for the charmonium system and a sum rule for the 
Adler function using the $V+A$ spectral function measured at LEP in
hadronic tau decays.
In contrast to the demonstration in the previous section, where within
our approximation, the vacuum was probed only by the temporal dynamics
of the heavy quarkonium system, the charmonium moments and the $V+A$
sum rules probe the full four-dimensional space-time
structure of the vacuum. We use the
$4$-dimensional version of the DOE with a common resolution scale 
for each coordinate as given in 
Eqs.\,(\ref{multDOEddim})--(\ref{g0000}). For convenience 
we define 
\begin{equation}
\Big\langle\frac{\alpha_s}{\pi}\,G^2\Big\rangle(\Omega)
\,\equiv\,
\frac{1}{2\pi^2}\,g_{0000}(\Omega)\,,
\label{runningG2}
\end{equation}
and in the following we will refer to it as the 
``running gluon condensate''. For $\Omega=\infty$ the
running gluon condensate coincides with the local gluon condensate 
$\langle(\alpha_s/\pi)G^a_{\rho\sigma}(0)G^a_{\rho\sigma}(0)\rangle$.

\subsection{Charmonium sum rules}
\label{subsectioncharmonium}

The determination of the local gluon condensate 
$\Big\langle\frac{\alpha_s}{\pi}\,G^2\Big\rangle(\infty)$ from
charmonium sum rules was pioneered in Refs.\,\cite{Novikov1}.
By now there is a vast literature on computations of Wilson
coefficients to various loop-orders and various dimensions of the 
power corrections in the
correlator of two heavy quark currents including updated
analyses of the corresponding sum rules (see
e.g. Refs.\,\cite{Nikolaev1,Nikolaev2}). Because the short-distance
coefficient of the running gluon condensate is $\Omega$-independent,
we can use the same strategy as in previous analyses, where the local
condensate was determined. The only difference is that we have 
to keep track of its dependence on the characteristic short-distance
scale.   

The relevant correlator is
\begin{equation}
\label{barccc}
(q_\mu q_\nu-q^2\,g_{\mu\nu}) \,\Pi^c(Q^2) 
\, = \, 
i\int d^4x\, \e^{iqx}\la T j^c_\mu(x)j^c_\nu(0)\ra
\,,\qquad
Q^2=-q^2
\,,
\end{equation}
where $j_\mu\equiv\bar{c}\gamma_\mu c$, and the $n$-th moment is
defined as 
\begin{equation}
{\cal M}_n
\, = \,
\frac{1}{n!}\,\left.\Big(\!-\frac{d}{dQ^2}\Big)^n
\,\Pi^c(Q^2)\,\right|_{Q^2=0}
\,.
\label{momdef1}
\end{equation}
Assuming analyticity of $\Pi^c$ in $Q^2$ 
away from the negative, real axis and employing 
the optical theorem, the $n$th moment can be expressed as a
dispersion integral over the charm pair cross section in $e^+e^-$
annihilation,
\begin{equation}
{\cal M}_n 
\, = \,
\frac{1}{12\pi^2\,Q_c^2}\,\int \frac{ds}{s^{n+1}}\,
\frac{\sigma_{e^+e^-\to c\bar
    c+X}(s)}{\sigma_{e^+e^-\to\mu^+\mu^-}(s)}
\,,
\label{momdef2}
\end{equation}
where $s$ is the square of the c.m.\,energy and $Q_c=2/3$.
We consider the ratio~\cite{Novikov1}
\begin{equation}
r_n 
\, \equiv \,
\frac{{\cal M}_n}{{\cal M}_{n-1}}
\label{mr}
\end{equation}
and extract the running gluon condensate as a function of $n$ from the
equality of the theore\-ti\-cal ratio using Eq.\,(\ref{momdef1}) and
the ratio based on Eq.\,(\ref{momdef2}) determined from experimental
data. 

The analytic form of the $n$-th moment in the local OPE up to
terms with dimension $6$ reads~\cite{Nikolaev1}
\begin{eqnarray}
{\cal M}_n 
& = &
\frac{3}{4\pi^2}\frac{2^n(n+1)(n-1)!}{(2n+3)!!}\,\frac{1}{(4m_c^2)^n}\,
\bigg\{\,
1 \, + \, \mbox{[pert. corrections]}
\nonumber\\[2mm]
 & & \hspace{1cm}
\,+\,
\delta^{(4)}_n\,\langle g^2 G^2\rangle 
\,+\,
\Big[\,
\delta^{(6)}_{G,n}\,\langle g^3 f G^3 \rangle
\,+\,
\delta^{(6)}_{j,n}\,\langle g^4 j^2 \rangle
\,\Big] \,+\,\ldots
\,\bigg\}
\,,
\nonumber\\[2mm]
\delta^{(4)}_n 
& = &
-\frac{(n+3)!}{(n-1)!\,(2n+5)}\,\frac{1}{9\,(4m_c^2)^2}
\,,
\nonumber\\[2mm]
\delta^{(6)}_{G,n} 
& = &
\frac{2}{45}\frac{(n+4)!\,(3n^2+8n-5)}{(n-1)!\,(2n+5)\,(2n+7)}\,
\frac{1}{9\,(4m_c^2)^3}
\,,
\nonumber\\[2mm]
\delta^{(6)}_{j,n} 
& = &
-\frac{8}{135}\,\frac{(n+2)!\,(n+4)\,(3n^3+47 n^2+244 n+405)}
      {(n-1)!\,(2n+5)\,(2n+7)}\,\frac{1}{9\,(4m_c^2)^3}
\,,
\label{Mexplicit}
\end{eqnarray}
where 
$\langle g^3 f G^3\rangle \equiv 
\langle g^3 f^{abc} G^a_{\mu\nu} G^b_{\nu\lambda}
G^c_{\lambda\mu}\rangle$
and
$\langle g^4 j^2 \rangle \equiv \langle g^4 j^a_\mu j^a_\mu\rangle$, 
$j^a_\mu$ being the light flavor singlet current.

In contrast to the calculation of the quarkonium energy level in
Sec.\,\ref{sectionquarkoniumlevel}, where we had a nonrelativistic
powercounting argument allowing for consistently considering only
two gluons interacting with the vacuum, the local OPE of the charmonium
moments is inconsistent if only the interaction of two gluons
with the vacuum is accounted for. This is because the Wilson
coefficients for condensates of dimension $6$ and higher are only
gauge-invariant if the interaction of any number of gluons that can
contribute at a certain dimension is taken into
account.\,\cite{Nikolaev1} Therefore, 
a reliable extraction of the running gluon condensate is only
possible if the higher dimensional local terms summed in 
$\delta^{(4)}_n\langle g^2 G^2\rangle(\Omega)$ represent a
reasonably good approximation to the full set of terms with the
corresponding dimensions in the local OPE. The most important
subleading contributions in the local OPE are the dimension $6$ terms
shown in Eq.\,(\ref{Mexplicit}). Using relation\,(\ref{2dG^2}), one can
derive that
\begin{equation}
\langle g^2 G D^2 G \rangle 
\, \equiv \,
\langle g^2 G_{\mu\nu}(0)\,D_\rho D_\rho\,G_{\mu\nu}(0)\rangle
\, = \,
2\,\langle g^4 j^2 \rangle \,- \,
2\,\langle g^3 f G^3 \rangle
\,.
\label{DDG2relation}
\end{equation}
Assuming further Eq.\,(\ref{pargfcsimple}), then the local dimension $6$
contribution contained in $\delta^{(4)}\langle g^2 G^2\rangle(\Omega)$
reads
\begin{equation}
\frac{1}{4\Omega^2}\,\delta^{(4)}_n\,
\langle g^2 G D^2 G \rangle 
\label{runningdim6est}
\end{equation}
In Tab.\,\ref{tabcharmonium} we have displayed, for $\Omega=2m_c/n$,
the ratio of the sum of local dimension $6$ contributions in the running gluon
condensate correction according to Eq.\,(\ref{runningdim6est}) and the 
sum of the local dimension $6$ term in Eq.\,(\ref{Mexplicit}) for $r_n$.  
We have used the standard values
$\langle g^3fG^2\rangle=0.045\,\,\mbox{GeV}^6$ obtained from the
instanton gas approximation~\cite{SVZ} and
$\langle g^4j^2\rangle=-\rho\, 4/3(4\pi)^2\alpha_s^2\langle \bar{q}q\rangle^2$
with $\alpha_s=0.7$ ($\mu\approx 0.7$~GeV) and 
$\langle \bar{q}q\rangle=-(0.24\,\,\mbox{GeV})^3$. The choice $\rho=1$  
corresponds to exact vacuum saturation. As explained later we
believe that $\Omega=2m_c/n$ is an appropriate choice for the
resolution scale.
\begin{table}[t!]  
\begin{center}
\begin{tabular}{|c|c|c|c|c|c|c|c|c|} \hline
$n$ & 1 &  2 &  3 &  4 &  5 &  6 &  7 &  8  
\\ \hline
$\rho=1$ & 0.13 & 0.36 & 0.56 & 0.73 & 0.87 & 0.99 & 1.08 & 1.16
\\ \hline
$\rho=2$ & 0.09 & 0.25 & 0.42 & 0.57 & 0.69 & 0.80 & 0.89 & 0.97
\\ \hline
\end{tabular}
\caption{Ratio of the local dimension-$6$ contributions contained in
the running gluon condensate and in the full OPE for $r_n$ as a
function of $n$.\label{tabcharmonium} 
}
\end{center}
\vskip 3mm
\end{table}
Indeed, one finds for the relevant ratios $r_n$ 
($2< n \lsim 7$) that the local dimension $6$ contribution  
contained in the running gluon condensate has the same sign and
roughly the same size as the dimension $6$ power correction 
in the full OPE for larger $n$, where higher dimensional contributions
are more important.

For the experimental moments we use the compilation presented in
Ref.\,\cite{Kuhn1}, where the spectral function is split into
contributions from the charmonium resonances, the charm threshold
region, and the continuum. For the latter the authors of
Ref.\,\cite{Kuhn1} used perturbation theory since no experimental data
are available for the continuum region. We have assigned a 10\% error
for the spectral function in the continuum region. 
We believe that this is sufficiently conservative 
since a large part of the error drops out in the ratio $r_n$.
For the purely perturbative contribution of the theoretical moments we
used the compilation of analytic ${\cal O}(\alpha_s^2)$ results from
Ref.\,\cite{Kuhn1} and adopted the $\overline{\mbox{MS}}$ mass
definition $\overline{m}_c(\overline{m}_c)$ (for any renormalization
scale $\mu$). For the Wilson coefficient of the gluon condensate we
have used the expression given in Eqs.\,(\ref{Mexplicit}).
We have checked that for $n\le 8$ the perturbative ${\cal
O}(\alpha_s^2)$ corrections do not exceed 50\% of the ${\cal
O}(\alpha_s)$ corrections for $\mu$ between $1$ and $4$~GeV.  

\begin{figure}[t!] 
\begin{center}
\leavevmode
\epsfxsize=3.5cm
\epsffile[260 430 420 720]{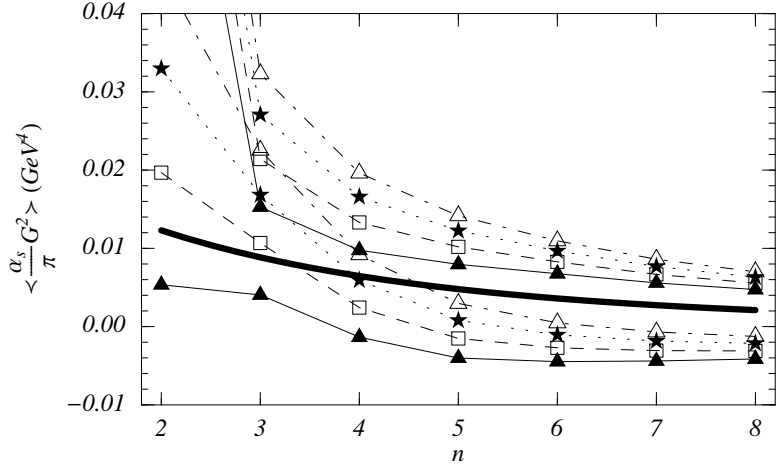}
%
%
\vspace{0cm}
 \caption{\label{figcharmonium} 
The running gluon condensate as a function of $n$ when extracted from
the ratio of moments $r_n$ for 
$\overline{m}_c(\overline{m}_c)=1.23$ (white triangles), $1.24$ 
(black stars), $1.25$ (white squares) and $1.26$~GeV (black
triangles). The area between the upper and lower symbols represents the
uncertainties. 
The thick solid line indicates the running gluon condensate as it is 
obtained from the
lattice-inspired ansatz in Eqs.\,(\ref{gfcfit})
}
 \end{center}
\end{figure}

Our result for the running gluon condensate as a function of
$n$ is shown in Fig.\,\ref{figcharmonium} for $n\le 8$ and 
$\overline{m}_c(\overline{m}_c)=1.23$ (white triangles), $1.24$ 
(black stars), $1.25$ (white squares) and $1.26$~GeV (black
triangles). For each value of $\overline{m}_c(\overline{m}_c)$ and $n$
the area between the upper and lower symbols represents the
uncertainty due to the experimental errors of the moments and variations of 
$\alpha_s(M_Z)$ between $0.116$ and $0.120$ and of $\mu$ between $1$
and $3$~GeV. 
We see that the running gluon condensate appears to be a decreasing
function of $n$. Let us compare this behavior of the running gluon
condensate with the one obtained from the lattice-inspired exponential
ansatz for $g(x)$ in Eq.\,(\ref{gfcfit}) using the definition of 
Eq.\,(\ref{gndefconvolution}),\footnote{
As explained in Sec.\,\ref{sectionanomdim},
Eq.\,(\ref{runningG2lattice}) corresponds to a computation of the
running gluon condensate in a regularization scheme with a cut-off
of order $\Omega$.
}
\begin{eqnarray}
\lefteqn{
\Big\langle\frac{\alpha_s}{\pi}G^2\Big\rangle_{\rm lat}(\Omega)
\, = \,
\frac{6\,A_0}{\pi^2}\,
\bigg\{\,
1 \, + \, \frac{1}{4\,\Omega^2\,\lambda_A^2} 
}
\nonumber\\[2mm]
& & \qquad
-\,
\frac{3\,\sqrt{\pi}}{4\,\Omega\,\lambda_A}\,
\bigg(1 + \frac{1}{6\,\Omega^2\,\lambda_A^2}\bigg)
\,\e^{1/(4\,\Omega^2\,\lambda_A^2)}
 \,\bigg(\,1 - {\rm erf}\Big(\frac{1}{2\,\Omega\,\lambda_A}\Big)
\,\bigg)
\,\bigg\}
\,.\qquad
\label{runningG2lattice}
\end{eqnarray}
Here, ${\rm erf}$ is the error function. Due to the exponential
behavior of the lattice-inspired ansatz for $g(x)$ the expression in
Eq.\,(\ref{runningG2lattice}) is a monotonic increasing function
of $\Omega$. Since the exact width of the
short-distance function $f$ for the moments is unknown we estimate it
using physical arguments. For large $n$, i.e.\,\,in the nonrelativistic
regime, the width is of the order of the quark c.m.\,kinetic energy 
$mv^2$, which scales like $m/n$ because the average quark velocity in
the $n$-th moment scales like $1/\sqrt{n}$.\,\cite{Hoang1}
For small $n$, on the other hand, the relevant short-distance scale
is just the quark mass. Therefore, we take $\Omega=2m_c/n$ as the most
appropriate choice of the resolution scale. In
Fig.\,\ref{figcharmonium} the running gluon condensate from
Eq.\,(\ref{runningG2lattice}) is displayed for
$\lambda_A^{-1}=0.7$~GeV and, exemplarily, $A_0=0.04$ (i.e. 
$\langle(\alpha_s/\pi)G^2\rangle_{\rm lat}
(\infty)=0.024\,\,\mbox{GeV}^4$)
and for $\Omega=(2.5\,\,\mbox{GeV})/n$ as the thick solid line. We
find that the $n$-dependence of the lattice inspired running gluon
condensate and our fit result from the charmonium sum rules are
consistent. This appears to confirm the lattice result for
the correlation length $\lambda_A$.
However, the uncertainties of our extraction are still quite
large, particularly for $n=2$, where the sensitivity to the error in
the continuum region is enhanced. 
We also note that the values for the running gluon condensate tend to 
decrease with increasing charm quark mass. For
$\overline{m}_c(\overline{m}_c)=1.27$~GeV we find that 
$\langle(\alpha_s/\pi)G^2\rangle$ ranges from $-0.01\,\,\mbox{GeV}^4$
to $+0.01\,\,\mbox{GeV}^4$ and shifts further towards negative values
for $\overline{m}_c(\overline{m}_c)\gsim 1.28$~GeV for all $n>2$.
Assuming the reliability of the DOE as well as the local OPE in
describing the nonperturbative effects in the charmonium sum rules
and that the Euclidean scalar functions $D(x)$ and $D_1(x)$
in Eqs.\,(\ref{pargfc})) are positive definite, our result disfavors 
$\overline{m}_c(\overline{m}_c)\gsim 1.28$~GeV.

\subsection{$V+A$ sum rule}

Recently, the spectral function for light quark production in the
$V+A$ channel has been remeasured by the 
Aleph~\cite{Aleph} and Opal~\cite{Opal} collaborations from hadronic
$\tau$ decays for $q^2\le m_\tau^2$. In the OPE the associated  
current correlator is dominated by the gluon condensate, and the 
dimension $6$ power corrections that are not due to a double
covariant derivative in the local gluon condensate are suppressed.  
The corresponding currents are
$j_\mu^{L/R}=\bar{u}\gamma_\mu(1\pm\gamma_5)d$ and the relevant
correlator in the chiral limit reads
\begin{equation}
i\int d^4\!x\, \e^{iqx}\la T j_\mu^{L}(x)j_\nu^{R}(0)\ra
\, = \,
(q_\mu q_\nu-q^2 g_{\mu\nu})\,\Pi^{\rm V+A}(Q^2)
\,,
\qquad
Q^2=-q^2
\,.
\label{VpAdef}
\end{equation}
Up to ${\cal O}(\alpha_s^3)$ the perturbative spectral function
is given as~\cite{Surguladze1} 
\begin{eqnarray}
\mbox{Im}\,\Pi^{\rm V+A}_{\rm pert}(-q^2-i\epsilon) & = & 
\frac{1}{2\pi}\,\bigg(\,
1 \, + \, \frac{\alpha_s(q^2)}{\pi} \, + \,
F_3\,\Big(\frac{\alpha_s(q^2)}{\pi}\Big)^2 \, + \,
F_4\,\Big(\frac{\alpha_s(q^2)}{\pi}\Big)^3
 +  \ldots
\bigg)
\,,
\nonumber\\[3mm]
F_3 
& = &
1.9857 - 0.1153\,n_\ell
\,,
\nonumber\\[3mm]
F_4 
& = &
-6.6368 - 1.2001\,n_\ell - 0.0052\,n^2_\ell
\,,
\label{VpAOPE}
\end{eqnarray}
where $n_\ell=3$ is the number of light quark flavors.  
The Wilson coefficient of the dimension-$4$ gluon condensate
correction is known to order $\alpha_s$,
\begin{equation}
\Pi^{\rm V+A}_{\rm np}(Q^2) 
\, = \,
\frac{1}{6\,Q^4}\,
\bigg(\,1-\frac{11}{18}\frac{\alpha_s}{\pi}\,\bigg)\,
\Big\langle\frac{\alpha_s}{\pi}G^2\Big\rangle
\, + \, \ldots
\,.
\label{VpAG2}
\end{equation} 
Since the correlator $\Pi(Q^2)^{\rm V+A}$ is cutoff-dependent itself,
we investigate the Adler function
\begin{equation}
\label{Adler}
D(Q^2) 
\, \equiv \, 
-Q^2\,\frac{\partial\,\Pi^{\rm V+A}(Q^2)}{\partial\, Q^2}
\, = \,
\frac{Q^2}{\pi}\,\int_0^\infty ds\, 
 \frac{\mbox{Im}\,\Pi^{\rm V+A}(s)}{(s+Q^2)^2}
\,.
\label{Adlerdef}
\end{equation}
For the corresponding experimental V+A spectral function we have used
the Aleph measurement~\cite{Aleph} in the resonance region up to 
$2.2$~GeV$^2$. For the continuum region above $2.2$~GeV$^2$ 
we used 3-loop perturbation theory for $\alpha_s(M_Z)=0.118$, and we
have set the renormalization scale $\mu$ to $Q$.
We note that the pion pole of the axial vector contribution has to be
taken into account in order to yield a consistent description in terms
of the OPE for asymptotically large $Q$.\,\cite{Braaten1}
We have checked that the known perturbative contributions to the Adler  
function show good convergence properties. For $Q$ between $1.0$ and
$2.0$~GeV and setting $\mu=Q$, the 3-loop (two-loop) corrections
amount to $5\%$ ($7\%$) and $0.5\%$ ($1.4\%$), respectively. 
On the other hand, the ${\cal O}(\alpha_s)$ correction to the
Wilson coefficient of the gluon condensate are between $-16\%$ and
$-6\%$ for $\mu$ between $0.8$ and $2.0$~GeV.
In analogy to the analysis for charmonium sum rules and setting $\Omega=Q$ 
we have compared the local dimension $6$ contributions contained in
the running gluon condensate in Eq.\,(\ref{VpAG2}) with the  
sum of all dimension $6$ terms in the OPE as they were determined in
Ref.~\cite{Braaten1}.  
The latter consist of local four-quark condensates, which we 
have estimated by assuming vacuum saturation at 
the scale $\mu\approx 0.7$ and using the parameters employed for the
charmonium sum rules. We again found that the corresponding 
dimension $6$ contributions have equal sign and roughly the same size. 

\begin{figure}[t!] 
\begin{center}
\leavevmode
\epsfxsize=3.5cm
\epsffile[260 430 420 720]{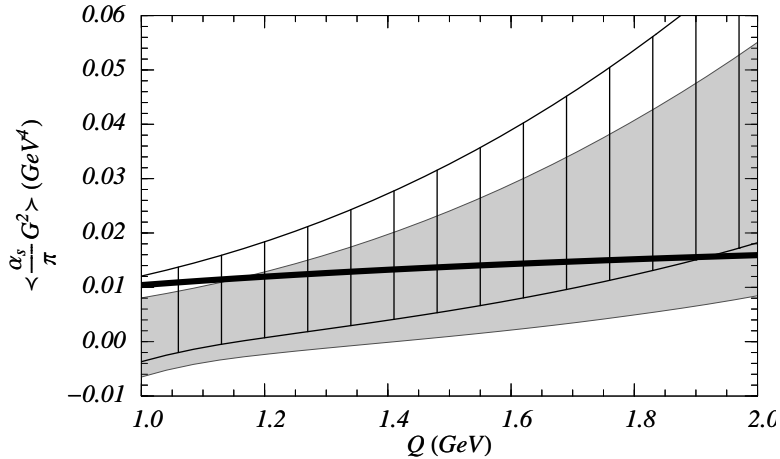}
%
%
\vspace{0cm}
 \caption{\label{figAdler} 
The running gluon condensate as a function of $Q$ when extracted from
the Adler function.
The grey area represents the allowed region using perturbation
theory at ${\cal O}(\alpha_s^3)$ and the striped region using
perturbation theory at  ${\cal O}(\alpha_s^2)$.
The thick solid line denotes the running gluon condensate from the
lattice-inspired ansatz in Eqs.\,(\ref{gfcfit}) for $\Omega=Q$.
}
 \end{center}
\end{figure}

Our result for the running gluon condensate as a function of $Q$ is
shown in Fig.\,\ref{figAdler}. 
for $1$\,GeV$\le Q \le 2$~GeV. The shaded region
represents the allowed range obtained from an analysis employing the
full ${\cal O}(\alpha_s^3)$ perturbative contribution and the striped 
region is obtained at ${\cal O}(\alpha_s^2)$. The uncertainties are due to 
the experimental errors in the spectral function and a variation of the
renormalization scale $\mu$ in the range $Q\pm 0.25$~GeV.
The strong coupling has been fixed at $\alpha_s(M_Z)=0.118$. 
We have restricted our analysis to the range 
$1\,\,\mbox{GeV}<Q<2\,\,\mbox{GeV}$ because for $Q\lsim
1$~GeV perturbation theory becomes unreliable and for $Q\gsim 2$~GeV the 
experimentally unknown part of the spectral 
function at $s\ge 2.2$\,GeV$^2$ is being probed.
The thick black line in Fig.\,\ref{figAdler} shows the
lattice-inspired running gluon condensate of 
Eq.\,(\ref{runningG2lattice}) for $\Omega=Q$ using the parameters from 
Sec.\,\ref{subsectioncharmonium}. 
We find that our result for the running gluon condensate as a function
of $Q$ is consistent with a function that increases with $Q$. 
The result is also consistent with the lattice-inspired expression
for the running gluon condensate, but the
uncertainties of our extraction are too large and the running of the
lattice-inspired condensate in the range between $1$ and $2$~GeV is
too small to draw a more definite conclusion.

\vspace{0.5cm}

\section{Anomalous dimensions}
\label{sectionanomdim}

In the previous sections we have presented the formalism for the
DOE assuming that the separation of fluctuations at
different length scales can be carried out without ambiguities and
that there is no need to introduce a regularization scheme. We have
applied the DOE for QCD quantities, where nonperturbative effects are
dominated by the gluon condensate, which can be defined such that its
anomalous dimension vanishes to all orders. This allowed us to ignore
anomalous dimensions at the level of precision intended in our
investigations in Sec.\,\ref{sectiongluoncondensate}.

However, in general, the separation of fluctuations at different
length scales is ambiguous and a regularization scheme is required to
carry out the separation systematically. In this
general case, the delocalized condensates cannot be defined through
the integrals in Eqs.\,(\ref{gndefconvolution}) or (\ref{fandgddim})
because the resolution scale $\Omega$ would itself serve as a
UV-regulator and, therefore, interfere with the regularization
scheme. Since the regularization of the short-distance coefficients in
the DOE remains 
unaffected, this would lead to inconsistencies in the DOE. Thus, in
the general case, the delocalized condensates are defined through the
sum over local condensates given in Eq.\,({\ref{fgrelations})
(and its $d$-dimensional generalization), which
should be considered more fundamental than
Eqs.\,(\ref{gndefconvolution}) and (\ref{fandgddim}). 
We emphasize that with this definition there is, by construction, no
interference between the DOE and the regularization scheme. 
Having in mind perturbative computations in quantum field theories in
general, this  means in particular that the delocalized expansion does
-- up to the summations that define the short-distance coefficients and
the matrix elements -- not affect renormalization, i.e.\,\,the 
structure of anomalous dimensions, the running of coefficients and
matrix elements, etc.. 

Coming back to the lattice-inspired expression for the running gluon
condensate in Eq.\,(\ref{runningG2lattice}) it is now clear that it
was determined in the exponential cutoff scheme defined through
Eq.\,(\ref{fandgddim}). However, because the
$\Omega$-dependence of the running gluon condensate is mainly due to
the exponential behavior of $g(x)$ and because the actual
renormalization-scale-dependence of the running gluon condensate is
small (due to the fact that it is generated by local condensates of
dimension $6$ and higher) the use of
expression\,(\ref{runningG2lattice}) for the comparisons in
Sec.\,\ref{sectiongluoncondensate}, where the perturbative
contributions were determined in the $\overline{\mbox{MS}}$ scheme,
are justified. 
 
It is obvious that the delocalized expansion can in principle be
applied to any quantum field theoretical computation, where a local
expansion can be carried out. In particular, it can be applied in
effective theories, and it allows for the systematic summation of
matrix elements of operators with additional covariant derivatives
in addition to the summation of
large logarithmic perturbative corrections obtained from solving the 
renormalization group equations for the short-distance
coefficients. In the following we briefly demonstrate the delocalized
expansion in the presence of a regularization scheme -- dimensional
regularization to be specific -- for $f_D/f_B$ at leading order in the
heavy quark mass expansion.
 
The meson decay constants are an important parameter in computations
of mixing and lifetime-differences. They can be computed in heavy
quark effective theory (HQET) in an
expansion in $\Lambda_{\rm QCD}/m_Q$, where $m_Q$ is the respective heavy
quark mass. To be definite, we consider the decay constant of a
pseudoscalar meson, which is defined as\footnote{
Since we discuss only the leading order in the $1/m_Q$ expansion, 
our treatment also applies to vector mesons.
}
\begin{equation}
\langle 0|\bar{q}\,\gamma^\mu\,\gamma^5\,Q|P(v)\rangle
\, = \,
i\,f_P\,\sqrt{M_P}\,v^\mu
\,,
\end{equation}
where $q$ and $Q$ are the light and heavy quark fields in full QCD,
$P(v)$ is the meson state with velocity $v$ in full QCD, and $M_P$ is
the mesons mass. In the (local) $1/m_Q$-expansion the current
$\bar{q}\gamma^\mu\gamma^5Q$ is written as a series of HQET currents
of increasing dimension each of which is multiplied by a Wilson coefficient:
\begin{equation}
\bar{q}\gamma^\mu \gamma^5 Q
\,\to \,
-\,C_0(m_Q,\mu)\,\bar{q}\gamma^5\gamma^\mu h_Q 
\, - \,
C_1(m_Q,\mu)\,\bar{q}\gamma^5 v^\mu h_Q
\, + \, \ldots
\,.
\end{equation}
Here, $h_Q$ denotes the heavy quark field in HQET and the ellipses
represent contributions from currents with higher dimensions. The
leading logarithmic expressions for the Wilson coefficients
read~\cite{Voloshin3}
\begin{equation}
C_0(m_Q,\mu) \, = \,
\bigg(\frac{\alpha_s(m_Q)}{\alpha_s(\mu)}\bigg)^{-\frac{2}{\beta_0}}
\,,
\qquad
C_1(m_Q,\mu) \, = \, 0
\,,
\label{WilsonC}
\end{equation}  
where $\beta_0$ is the one-loop beta-function. Since the
vacuum-to-meson matrix element of the current
$\bar{q}\gamma^5\gamma^\mu h_Q$ 
is equal for $B$ and $D$ mesons at leading order in $1/m_Q$, this
leads to the leading order relation
\begin{equation}
\frac{f_D}{f_B} \, = \,
\sqrt{\frac{M_B}{M_D}}\,
\bigg(\frac{\alpha_s(m_b)}{\alpha_s(m_c)}\bigg)^{\frac{6}{25}}
\label{fDufBHQET}
\end{equation}
for $4$ active light flavors.
Recent lattice measurements~\cite{latticefDfB} indicate that 
\begin{equation}
\frac{f_D}{f_B} 
\, \simeq \,
\mbox{1.0 -- 1.3}
\,,
\end{equation}
but the RHS of Eq.\,(\ref{fDufBHQET}) gives 
$f_D/f_B\approx 1.5$. It was found in Ref.~\cite{Neubert1} that the 
discrepancy cannot be accounted for quantitatively by subleading
$1/m_Q$ contributions, because the corresponding corrections are
simply too large to allow for a quantitative improvement of the leading
order relation in Eq.\,(\ref{fDufBHQET}). In particular, for the $D$
meson the local $1/m_Q$ expansion seems to break down, and it
appears that going beyond the leading order approximation for
$f_D/f_B$ is meaningless.

We suspect that the same statement might also apply to the
delocalized version of the $1/m_Q$ expansion, but from the
examinations in the previous parts of this paper one can expect that
the delocalized expansion might provide a better leading order
approximation for an appropriate choice of resolution scales. In the
delocalized $1/m_Q$ expansion the leading order short-distance
coefficients are equivalent to the local ones shown in
Eq.\,(\ref{WilsonC}). However, the vacuum-to-meson matrix element of
the current $\bar{q}\gamma^5\gamma^\mu h_Q$ becomes a
resolution-scale-dependent quantity because it sums matrix elements of 
local currents of the generic form $D^{2n}\bar{q}\gamma^5\gamma^\mu h_Q$
($n\ge 0$). Evidently, the proper choice of the resolution scale is
$\Omega\sim m_Q$.
Thus the leading order expression for $f_D/f_B$ in the
delocalized expansion can be written as
\begin{equation}
\frac{f_D}{f_B} \, = \,
\sqrt{\frac{M_B}{M_D}}\,
\bigg(\frac{\alpha_s(m_b)}{\alpha_s(m_c)}\bigg)^{\frac{6}{25}}\,\rho
\,,
\label{fDufBDOE}
\end{equation}
where
\begin{equation}
\rho 
\, = \,
\frac{\langle 0|\bar{q}\,\gamma^5\gamma^\mu\,
h_c|D(v)\rangle(\mu,\Omega=m_c)}
       {\langle
	 0|\bar{q}\,\gamma^5\gamma^\mu\,
h_b|B(v)\rangle(\mu,\Omega=m_b)}
\,.
\label{ratioDOE}
\end{equation}
The common renormalization scale $\mu$ in the $\Omega$-dependent matrix
elements is set below the charm quark mass, but not fixed otherwise.
Comparing expression~(\ref{fDufBDOE}) with the lattice result one finds
$\rho\simeq\mbox{0.6--0.8}$. (Neglecting the anomalous dimension of the
current $\bar{q}\gamma^5\gamma^\mu h_Q$ gives essentially the same
result due to the large uncertainty of the lattice measurements.) We
are not aware of any other independent determination of the ratio
$\rho$, but this scaling relation has an interesting physical
interpretation in the framework of a model, where one assumes that the
$B$ and the $D$ meson wave functions have approximately the same
size\footnote{  
The results from the lattice simulations of Ref.~\cite{DeGrand1}
indicate that this assumption is not unrealistic. This assumption is
also consistent with results from potential model
computations.\,\cite{Hwang1} 
}
and that the vacuum-to-meson matrix elements of the local currents 
$D^{2n}\bar{q}\gamma^5\gamma^\mu h_Q$ scale like inverse powers of the meson
size. 
Assuming exemplarily that the meson wave functions 
have a Gaussian drop-off behavior (see e.g. Ref.\,\cite{Hwang1})
$\langle 0|\bar{q}(-\bmx/2)\gamma^5\gamma^\mu\,S(-\bmx/2,\bmx/2)\,
h_Q(\bmx/2)|P\rangle\sim\exp(-|\bmx|^2/\lambda^2)$ 
one finds, using the $3$-dimensional version of
Eq.\,(\ref{fandgddim}), 
\begin{equation}
\rho 
\, \simeq \,
\left(\,\frac{1+\frac{1}{\lambda^2 m_b^2}}{1+\frac{1}{\lambda^2 m_c^2}}
\,\right)^{\frac{3}{2}}
\,.
\end{equation}
One can then derive that $\lambda^{-1}\simeq\mbox{0.5--0.8}$~GeV,
which gives an estimate for the meson size. 
On the other hand, for an exponential drop-off behavior
$\sim\exp(-|\bmx|/\lambda)$ one obtains $\lambda^{-1}\simeq\mbox{0.3--0.7}$~GeV.

\vspace{0.5cm}

\section{Summary and outlook}
\label{sectionsummary}

In this paper we have proposed a new type of short-distance expansion
of correlators of gauge invariant currents in QCD, which we name
delocalized operator expansion (DOE). This expansion
originates from nonlocal projections of gauge invariant correlation
functions based on a delocalized version of the multipole expansion
for the perturbatively calculable coefficient functions. Whereas the
usual (local) OPE is based on a multipole expansion where the
short-distance process in configuration space is written as the sum of
delta-functions and derivatives of delta-functions, the DOE is based
on a delocalized version of the multipole expansion using functions
that have the width $\Omega^{-1}$. We call $\Omega$ the resolution 
parameter. In this paper we have constructed such a delocalized
multipole expansion based on the Gaussian function and using the
orthonormality and completeness properties of the  Hermite
polynomials. We emphasize that other, similar constructions are
possible which have properties that are comparable to the one presented
here and that they might be even better suited to certain problems than the
one we have used here. 

In the DOE the condensates become
quantities that are dependent on the resolution parameter $\Omega$ and
are related to an {\sl infinite} sum of VEV's of local operators with
equal and larger number of covariant derivatives. The short-distance
coefficients also become dependent on the resolution parameter
$\Omega$ and are related to a {\sl finite} sum of short-distance
coefficients to local operators with equal and fewer number of
covariant derivatives. The relative weight of the local terms in these
sums is governed by the resolution parameter $\Omega$.  
For $\Omega\to\infty$ the DOE reduces to the usual local OPE.
As is the local OPE, the DOE is parametrically an expansion in powers 
of $\Lambda/Q$, where $\Lambda$ denotes the highest nonperturbative
scale occurring in the  corresponding gauge invariant correlation
function and $Q$ is the external short-distance scale at which the
vacuum is being probed. If the resolution scale is chosen to
be of order $Q$, there is an additional suppression of higher orders
in the expansion by powers of a small number.  This can be understood
intuitively, because for $\Omega\sim Q$ the leading term in the
delocalized multipole expansion represents a better approximation of
the short-distance process than the delta-function. In this way the
DOE can account for non-local effects.

Calculating the nonperturbative correction to the perturbatively 
determined ground state energy of a heavy quarkonium system, 
we have demonstrated the improved convergence properties of the 
delocalized expansion using a lattice-inspired toy model for the 
gluonic $2$-point field strength correlator. Reversing the situation,
we have extracted the running gluon condensate
$\langle(\alpha_s/\pi)G^2\rangle(\Omega)$ from experimental data  
using charmonium sum rules and the Adler function in the 
$V+A$ channel of light quark pair production. We found strong evidence
that the gluon condensate is indeed a resolution-dependent quantity
and our results are consistent with recent lattice measurements of the 
correlation length of the gluon field strength correlator. 

The DOE can be applied in the framework of effective theories.
By construction, the DOE does not interfere with the regularization
scheme since the resolution-dependent short-distance coefficients and
matrix elements are defined as sums of local short-distance
coefficients and matrix elements, respectively. Thus, the DOE does not
affect the renormalization properties of a theory. In the DOE we have 
derived the leading order expression for $f_D/f_B$ in the heavy quark
mass expansion.

Further investigations and more detailed applications of the DOE
are planned in forthcoming work.

\vspace{1cm}

\begin{appendix}

\section{Tensor structure of $\langle g^2 G D^2 G \rangle$}
\label{appendixtensor}

Here we show that the lattice-implied dominance of the
tensor structure in Eq.\,(\ref{pargfcsimple}) for the gluon field
strength correlator is consistent with a phenomenological
determination of the dimension 6 condensate
$\langle g^2 G_{\mu\nu}(0) D_{\alpha}D_{\beta}
G_{\kappa\lambda}(0)\rangle$.
When evaluated in Schwinger gauge $x_\mu A_\mu=0$~\cite{Schwinger1} 
we are, according to Eq.\,(\ref{g0000}), 
concerned with the quantity $\la g^2 G^a_{\mu\nu}\pd_\rho\pd_\rho
G^a_{\alpha\beta}\ra$   
taken at $x=0$. Following Ref.\,\cite{Nikolaev1} it can be
parametrized  in Euclidean space-time as
\begin{eqnarray}
\label{2dG^2}
-\la g^2 G^a_{\mu\nu}\pd_\rho\pd_\rho G^a_{\alpha\beta}\ra 
& = & 
8\,O^-\left(\delta_{\mu\beta}\delta_{\alpha\nu}-
\delta_{\mu\alpha}\delta_{\nu\beta}\right)
\nonumber\\[2mm]    
& & 
+ \,O^-\left(\delta_{\mu\beta}\delta_{\alpha\nu}+
\delta_{\alpha\nu}\delta_{\mu\beta}-\delta_{\alpha\mu}\delta_{\nu\beta}-
\delta_{\beta\nu}\delta_{\mu\alpha}\right)
\nonumber\\[2mm] 
& &
+\,O^+\left(\delta_{\mu\beta}\delta_{\alpha\nu}+
\delta_{\mu\beta}\delta_{\alpha\nu}- 
\delta_{\mu\alpha}\delta_{\nu\beta}-
\delta_{\nu\beta}\delta_{\mu\alpha}\right)
\,,
\end{eqnarray}
where
\begin{equation}
\label{O+-}
O^{\pm}
\, \equiv \,
\frac{1}{72}\la g^4 j^a_\mu j^a_\mu\ra\pm
\frac{1}{48}\la g^3f_{abc}G^a_{\mu\nu}G^b_{\nu\lambda}G^c_{\lambda_\mu}\ra\,,
\end{equation}
and $j^a_\mu$ denotes a light flavor-singlet current. In Eq.\,(\ref{2dG^2})  
the tensor structure multiplying $8\,O^-$ is the same as the one
resulting from an omission of the terms involving  $D_1$ in the
parametrization of Eq.\,(\ref{pargfc}). Now, we consider
\begin{equation}
\label{contr}
\la g^2 G^a_{\mu\nu}\pd_\rho\pd_\rho G^a_{\mu\nu}\ra
\, =\, 
96\,Q^-+24\,(O^-+O^+)\,,
\end{equation}
and we have to show that, phenomenologically, the 
contribution $96\,Q^-$ is the dominant one, that is
\begin{equation}
\label{RO}
R_O \, \equiv \, \frac{96\,Q^-}{24\,(O^-+O^+)}
\, \gg \, 
1\,. 
\end{equation}
Assuming exact vacuum saturation to express $\la g^4 j^2\ra$ 
in terms of a square of the 
quark condensate and using $\alpha_s=0.7$ ($\mu=0.7$~GeV), 
$\la g^3 f_{abc}G^a_{\mu\nu}G^b_{\nu\lambda}G^c_{\lambda\mu}\ra=0.045$\,GeV$^6$, and 
$\langle \bar{q}q\rangle=-(0.24\,\,\mbox{GeV})^3$ as in Sec.\,5.1, we obtain
\begin{equation}
\frac{\la g^3 f_{abc}G^a_{\mu\nu}G^b_{\nu\lambda}G^c_{\lambda\mu}\ra}
{\la g^4 j^a_\mu j^a_\mu\ra}
\, \sim \, -2.3\,.
\end{equation}
For the ratio $R_O$ of Eq.\,(\ref{RO}) this implies 
\begin{equation}
R_O \, \sim \, 8.8
\,,
\end{equation}
which justifies the omission of the terms in 
the second and third line of Eq.\,(\ref{2dG^2}).

\section{Short-distance coefficients for quarkonium ground states energy
    levels} 
\label{appendixshort}

\begin{eqnarray}
f_0(\infty) & = & \frac{702}{425}\,\frac{m}{36\,k^4}
\,,
\\[2mm] 
f_2(\infty) & = & \frac{1108687904}{844421875}\,\frac{m^3}{36\,k^8}
\,, 
\\[2mm]
f_4(\infty) & = & \frac{17608502062304825344}{2270010244541015625}\,
                  \frac{m^5}{36\,k^{12}}
\,,    
\\[2mm]
f_6(\infty) & = & \frac{45707850430207738594209974779904}
                        {350274120613508115758056640625}\,
                  \frac{m^7}{36\,k^{16}}
\,, 
\\[2mm]
f_8(\infty) & = & \frac{
         3377025742042031485195560438171693564080529866752}
        {754795366909616960779550333930811309814453125}\,
                  \frac{m^9}{36\,k^{20}}
\,.\quad\mbox{}
\end{eqnarray}

\end{appendix}

\vspace{1cm} 

\section*{Acknowledgements}

The authors would like to thank V. I. Zakharov 
for helpful comments and discussions. We also acknowledge 
useful conversations with U.~Nierste, R.~Sommer and H.~Wittig.

\vspace{1cm}

\bibliographystyle{prsty}

\end{document}